\documentclass[fleqn,usenatbib]{mnras}

\usepackage{newtxtext,newtxmath}

\usepackage[T1]{fontenc}
\usepackage{ae,aecompl}

\usepackage{graphicx}	
\usepackage{amsmath}	
\usepackage{diagbox}
\usepackage{pifont}
\usepackage{float}
\title[Striped jet model]{Connecting the early afterglow to the prompt GRB and the central engine in the striped jet model}

\author[Michail Damoulakis et al.]{
Michail Damoulakis,$^{1}$\thanks{E-mail: mdamoula@purdue.edu}
Rodolfo Barniol Duran,$^{2}$
Dimitrios Giannios$^{1}$
\\
$^{1}$Department of Physics, Purdue University, 525 Northwestern Avenue, West Lafayette, IN 47907 USA\\
$^{2}$Department of Physics and Astronomy, California State University, Sacramento, 6000 J Street, Sacramento CA 95819-6041 USA
}

\date{Accepted XXX. Received YYY; in original form ZZZ}

\pubyear{2022}

\begin{document}
\label{firstpage}
\pagerange{\pageref{firstpage}--\pageref{lastpage}}
\maketitle

\begin{abstract}
Despite a generally accepted framework for describing the Gamma-Ray Burst (GRB) afterglows, the nature of the compact object at the central engine and the mechanism behind the prompt emission remain debated. The striped jet model is a promising venue to connect the various GRB stages since it gives a robust prediction for the relation of jet bulk acceleration, magnetization and dissipation profile as a function of distance. Here, we use the constraints of the magnetization and bulk Lorentz of the jet flow at the large scales where the jet starts interacting with the ambient gas in a large sample of bursts to (i) test the striped jet model for the GRB flow and (ii) study its predictions for the prompt emission and the constraints on the nature of the central engine. We find that the peak of the photospheric component of the emission predicted by the model is in agreement with the observed prompt emission spectra in the majority of the bursts in our sample, with a radiative efficiency of about 10 per cent. Furthermore, we adopt two different approaches to correlate the peak energies of the bursts with the type of central engine to find that more bursts are compatible with a neutron star central engine compared to a black hole one. Lastly, we conclude that the model favors broader distribution of stripe length-scales which results in a more gradual dissipation profile in comparison to the case where the jet stripes are characterized by a single length-scale. 

\end{abstract}




\section{Introduction}
 Our understanding of the gamma-ray burst (GRB) prompt emission has greatly improved in the last decades thanks to the contributions of space-based Gamma-ray observatories such as the {\it Swift} and {\it Fermi} satellites (see, e.g., \citealp{Bosnjak2022}). Ground and space-based observatories have also contributed in the study of the afterglow emission and provide insight into the interactions of the jet with the interstellar medium (see, e.g., \citealp{KumarZhang2015}).  
 
Over the past 10-15 years, multi-wavelength efforts have achieved very fast response times (down to seconds) to GRB triggers. Such efforts have allowed us to supplement information on the early stages of the jet interactions with the ambient gas. The emission from these stages (afterglow) carries invaluable clues for the internal properties of the jet flow, such as the bulk Lorentz factor $\Gamma$ and the jet magnetization $\sigma$ (defined as the ratio between magnetic energy density and enthalpy) at a large distance from the central engine. For instance, the time of the peak of the external shock emission marks the onset of the jet deceleration and, therefore, depends sensitively on the jet bulk $\Gamma$ (e.g., \citealp{SariPiran1998,PanaitescuKumar2000}). Moreover, the initial deceleration of the jet is driven by a reverse shock into it (e.g., \citealp{GranotTaylor2005,ZhangKobayashi2005,McMahon2006,MimicaGiannios2009}). The emission from the reverse shock, typically peaking at optical wavelengths, has been used to infer the magnetic field strength of the jet flow (e.g., \citealp{Gomboc2008,Japelj2014})

However, despite decades of theoretical research and observations, and a general framework for describing the GRB afterglows, the nature of the compact object (black hole or neutron star) and the jet properties remain debated. There are several aspects of the prompt emission that are still highly uncertain; the high variability of their light curve and the plethora of differences observed from burst to burst makes it difficult to provide a robust theoretical model that describes all of the following: What is the composition of the jet flow and the mechanisms behind the prompt phase? Are the jets thermally or magnetically dominated? What mechanism is responsible for the bulk acceleration of the ejecta and the particle acceleration responsible for the observed emission? Does the dissipation happen predominantly below or above the Thomson photosphere of the jet? 

The "striped jet" model is a promising venue to investigate many of these questions (\citealp{Spruit2001,Drenkhahn2002,DrenkhahnSpruit2002,GianniosSpruit2005,McKinneyUzdensky2012, Giannios_Uzdensky2019}). In this model, a magnetically dominated jet is endowed with small-scale field reversals or "stripes", due to the field's non-axisymmetry at the central engine. Dissipation in the stripes, due to magnetic reconnection, drives both the jet bulk acceleration and particle energization out to a distance where most of the magnetic energy has been used up and the acceleration ceases. The model gives a robust prediction on the relation of jet acceleration and magnetization and their dependence as a function of distance. The simplest striped jet model has a single characteristic stripe length $l_{\rm min}$ \citep{Drenkhahn2002}; however this model has recently been generalized by \cite{Giannios_Uzdensky2019} to consider a distribution of stripe lengths scaling as $l^{-\alpha}$ for $l>l_{\rm min}$ and $\alpha>1$. Critical parameters of the model are the asymptotic Lorentz factor $\Gamma_\infty$ and the stripe parameter $\alpha$  \citep{Giannios_Uzdensky2019}. Dissipation in this model takes place mostly at compact scales, below and around the Thomson photosphere for slower flows, and at optically thin conditions for faster ones. The prompt emission spectra predicted by the model have been investigated in detail in both limits and can be used for direct comparison with prompt observations (e.g., \citealp{Giannios2008, Giannios_Uzdensky2019, Gill2020}).  

In this paper, we use the deceleration radius of the GRB blast wave to provide the critical connection between the afterglow and the prompt emission stages of GRBs. Using results of afterglow modeling, we can infer the values of the bulk Lorentz factor and the magnetization at the deceleration radius. These values can be used as the outer boundary conditions that the striped jet model must satisfy, allowing us to move closer to the GRB central engine, where observations are scarce, and obtain valuable information on the nature of the central engine. The model also provides the location in the jet where most of the energy is dissipated, the location of the Thomson photosphere, the expected photospheric peak energy and the photospheric radiative efficiency. All these quantities can be compared to the available prompt-phase observations.   

The paper is organized as follows. In section 2, we present the theory of the striped jet model for the GRB flow where we describe how the acceleration profile, as well as the magnetization of the jet, varies for different values of the stripe parameter $\alpha$. In section 3, we calculate the power of the thermal (photospheric) and non-thermal components of the jet, and use them to calculate the corresponding radiative efficiencies, as well as the peak energy of the photopsheric component. In section 4, we apply the theory to two different samples of GRBs where crucial jet properties are determined by the afterglow modeling. Finally, in section 5 we summarize our results, discuss the significance of the striped jet model, and propose ways to improve our theoretical modeling in future work. 

\section{Theory of the Striped Jet Model}
 Here, we first summarize the dynamics of the \cite{Drenkhahn2002} model (see also \citealp{DrenkhahnSpruit2002}), where the jet is injected with a single length-scale for the magnetic field reversals (single stripe-width model). Then we extend the discussion to the multiple-stripe model, as developed by \cite{Giannios_Uzdensky2019}. From now on, primed (')  and unprimed quantities correspond to the comoving and laboratory frame, respectively.

\subsection{Dissipation on a single scale}
In the \cite{Drenkhahn2002} (D02) model, the jet contains stripes where the magnetic field reverses sign at a typical length scale $l'_{\text{min}}$. Between stripes,  dissipation takes place via magnetic reconnection at a speed which is a fraction $\epsilon$ of the comoving Alfv\'en velocity $v'_{\text{A}}$. The dissipation timescale in the comoving frame in this case is
\begin{equation}
 \tau'_{\text{min}}=\frac{l'_{\text{min}}}{u'_{\text{rec}}}=\frac{l'_{\text{min}}}{\epsilon v'_{\text{A}}}.
 \label{eq:taumin}
\end{equation}
For the rest of the paper, we assume $\epsilon\simeq 0.1$ \citep{Drenkhahn2002}, that is, the magnetic field reconnects with a speed of approximately 10 per cent of the Alfv\'en velocity. Also, if the flow is Poynting flux-dominated, then $v'_{\text{A}}\approx c$. Assuming an ultra relativistic flow, the approximate velocity profile of the striped jet is
\begin{equation}
     \Gamma(r)=\begin{cases}
    \Gamma_{\infty}\left(\dfrac{r}{R_{\text{diss}}} \right)^{1/3}, & r\leq R_{\text{diss}}\\
    \Gamma_{\infty}, & r>R_{\text{diss}},
  \end{cases}
\end{equation}
where the dissipation radius is defined by
\begin{equation}
    \label{eq:rdiss}
    R_{\text{diss}}=R_0\Gamma_{\infty}^2.
\end{equation}
 $R_{\text{diss}}$ is the distance where the dissipation due to magnetic reconnection peaks - as defined in \cite{Drenkhahn2002}, and $R_0\equiv\frac{l_{\text{min}}}{6\epsilon}$ is a characteristic scale of the model that, as we discuss in the next section, depends strongly on the nature of the central engine (black hole or neutron star). 

An important parameter of the model is the total luminosity of the jet $L$, which is the sum of the kinetic and Poynting components, and it is given by
\begin{equation}
    L=L_k+L_p=\Gamma \dot{M} c^2 (1+\sigma),
\end{equation}
where we defined $\sigma(r) \equiv \frac{L_p}{L_k}$ as the magnetization of the jet at a distance $r$ from the central engine.

Our model assumes that all the Poynting flux dissipates at asymptotically large distance (i.e., there is no non-reconnecting component of the magnetic field). Using conservation of energy, and the assumption that the magnetization drops to $\sigma\rightarrow 0$ as $r\rightarrow \infty$, we obtain
\begin{equation}
\label{eq:gammasigma}
    \Gamma_{\infty}=\Gamma(r)[1+\sigma(r)].
\end{equation}

\subsection{Estimates for $R_0$} \label{section:R0}

The characteristic scale of the model $R_0= l_{\rm min}/6\epsilon$ depends on the nature of the central engine and its immediate surroundings. For a neutron star as the central engine, where the magnetic field axis is not aligned to the spin axis, the length scale $l_{\text{min}}$ over which the magnetic field reverses sign is closely connected to the rotation period of the star, that is, $l_{\text{min}}=\frac{2\pi c}{\Omega}$. The characteristic scale will then be $R_0=\frac{\pi c}{3\epsilon\Omega}$. The range of $R_0$ for a neutron star engines is determined by the range of the initial rotation frequencies. Neutron stars born with initial sub-ms period loose energy fast through gravitational wave emission while slow $\gtrsim 10$~ms rotators do not have sufficient rotational energy to power the observed GRBs. We can, therefore, assume that $\frac{2\pi}{\Omega}\approx$1-10~ms for typical NS rotation periods (hence, the "millisecond magnetar" model, see, e.g. \citealp{Duncan1992,Metzger2011,Kasen2010, Margalit2018}). In that case, we expect roughly $R_{\text{0,NS}}\approx(6\times10^7-6\times10^8\:)$cm. 

For a black hole central engine, the range of $l_{\text{min}}$ is related to the characteristic timescale that magnetic polarity flips are generated at the surrounding accretion disk and advected into the black hole. Such flips in the polarity of the magnetic field through the black hole have been observed in general relativistic MHD simulations of accretion tori around a black hole (\citealp{Christie2019}). Here we estimate that the flip timescale is about 10 times longer than the Keplerian period of the disk at the location where the disk dissipation peaks (see, e.g., \citealp{Giannios_Uzdensky2019}).

More specifically, assuming that the accretion disk surrounding the black hole is described by the General Relativistic model of \cite{ThorneNovikov1973}, the distance of peak emission $\rho_{\text{L,max}}$ depends only on the mass of the black hole, and the dimensionless spin parameter $a_{*}$ (not to be confused with the stripe parameter $\alpha$ discussed in detail in the next subsection). Thus, assuming a timescale of 10 Keplerian periods at that distance, we obtain $l_{\text{min}}\approx 10T_K(\rho_{\text{L,max}})c=20\pi\sqrt{c^2/GM}\rho_{\text{L,max}}^{3/2}$. A typical range of the spin parameter is $a_{*}\approx0.7-0.98$, and for a range of about 7-15 $M_{\odot}$ black hole (\citealp{TchekhovskoyGiannios2015}), we obtain roughly $R_{\text{0,BH}}\approx(6\times 10^8-6\times 10^{9})$ cm. 

To summarize, the parameter $R_0$ is highly uncertain and model dependent, since different assumptions about the central engines (e.g. magnetar frequency and black hole mass, the existence or not of an accretion disk) will affect the outcome.  
Nonetheless, we can use these estimates as a guideline for this work, where a neutron star central engine has values of $R_0$ that are one order of magnitude smaller than the ones for a black hole. 

\subsection{Dissipation on multiple scales}
In a generalization of the D02 model, the magnetic field can dissipate on multiple length scales that follow a power-law distribution given by
\begin{equation}
     \mathcal{P}_{\alpha}(l)=\frac{\alpha-1}{l_{\text{min}}}\left(\frac{l}{l_{\text{min}}} \right)^{-\alpha},
\end{equation}
where $l_{\text{min}}$ corresponds to the minimum spatial scale of field reversal and $\alpha$ to the stripe parameter, which determines the broadness of the stripe distribution. 

The way this distribution of stripes is defined is not unique, but the power-law assumption is a choice based on physical arguments, both for a neutron star and a black hole central engine. We expect the stripes formed from a neutron star to be strongly dependent on its rotation period (stripes having a dominant characteristic length $l_{\text{min}}$), while for an central engine containing an accretion disk around a black hole or a neutron star we may expect a stripe distribution $\sim l^{-5/3}$ (\cite{Giannios_Uzdensky2019}). Both scenarios can be considered special cases of a power-law distribution $\sim l^{-\alpha}$. We note here that on this work we focus on stripe distributions with $\alpha>1$, which means that the dissipation power is dominated by the small stripes $l\gtrsim l_{\text{min}}$. Using equation (\ref{eq:taumin}) we find that for stripes of width $l$, the radius along the jet at which they dissipate is $r_{\text{rec}}=ct_{\text{rec}}\simeq \frac{\Gamma^2}{\epsilon}l$, where $t_{\text{rec}}=\Gamma t'_{\text{rec}}$ and $t'_{\text{rec}}=l'/u'_{\text{rec}}\simeq \Gamma l/(\epsilon c)$. As a result, the smaller stripes will dissipate first. At larger distances, only the stripes with large characteristic length scales remain. 

In this model, the velocity profile of the jet is given by \cite{Giannios_Uzdensky2019}:
\begin{equation}
    \frac{d\Gamma}{dr}=\dfrac{1}{3R_{\text{diss}}}\dfrac{(\Gamma_{\infty}-\Gamma)^{\frac{1-3\alpha}{2-2\alpha}}}
    {\Gamma_{\infty}^{\frac{3\alpha-5}{2-2\alpha}}\Gamma^2}.
    \label{eq:dGammadr}
\end{equation}
Changing the variables, $\chi\equiv\frac{\Gamma}{\Gamma_{\infty}}, \zeta\equiv\frac{r}{3R_{\text{diss}}},k\equiv\frac{1-3\alpha}{2-2\alpha}$ we obtain the dimensionless form of equation (\ref{eq:dGammadr}), which is given by
\begin{equation}
    \frac{d\chi}{d\zeta}=\frac{(1-\chi)^k}{\chi^2}.
    \label{eq:xzeq}
\end{equation}

Here, we keep the same definition for $R_{\text{diss}}$ from D02 since it corresponds to the distance from the central engine where magnetic dissipation peaks, independently from the value of $\alpha$. The difference between small and large values of $\alpha$ appears only in the broadness of the dissipation profile (see , e.g., \citealp{Giannios_Uzdensky2019}).  

Solving equation (\ref{eq:xzeq}), we obtain the dimensionless distance $\zeta$ as a function of the normalized Lorentz factor $\chi$ for different values of the parameter $k$ as follows
\begin{equation}
    \label{eq:zetachi}
     \zeta_k(\chi)=\begin{cases}
    \frac{(1-\chi)^{3-k}}{k-3}-\frac{2(1-\chi)^{2-k}}{k-2}+\frac{(1-\chi)^{1-k}}{k-1}+C_k, & k\neq\{1,2,3\}.\\
    2(1-\chi)-\frac{(1-\chi)^2}{2}-\ln(1-\chi)-\frac{3}{2}, & k=1. \\
    \frac{1}{1-\chi}+\chi+2\ln(1-\chi)-1   & k=2\\
    \frac{1}{2(1-\chi)^2}-\frac{2}{1-\chi}-\ln(1-\chi)+\frac{3}{2}   & k=3,    
  \end{cases}
\end{equation}
where $C_k=-(k-3)^{-1}+2(k-2)^{-1}-(k-1)^{-1}$ (\citealp{Giannios_Uzdensky2019}).
 
\begin{figure}
	\includegraphics[width=\columnwidth]{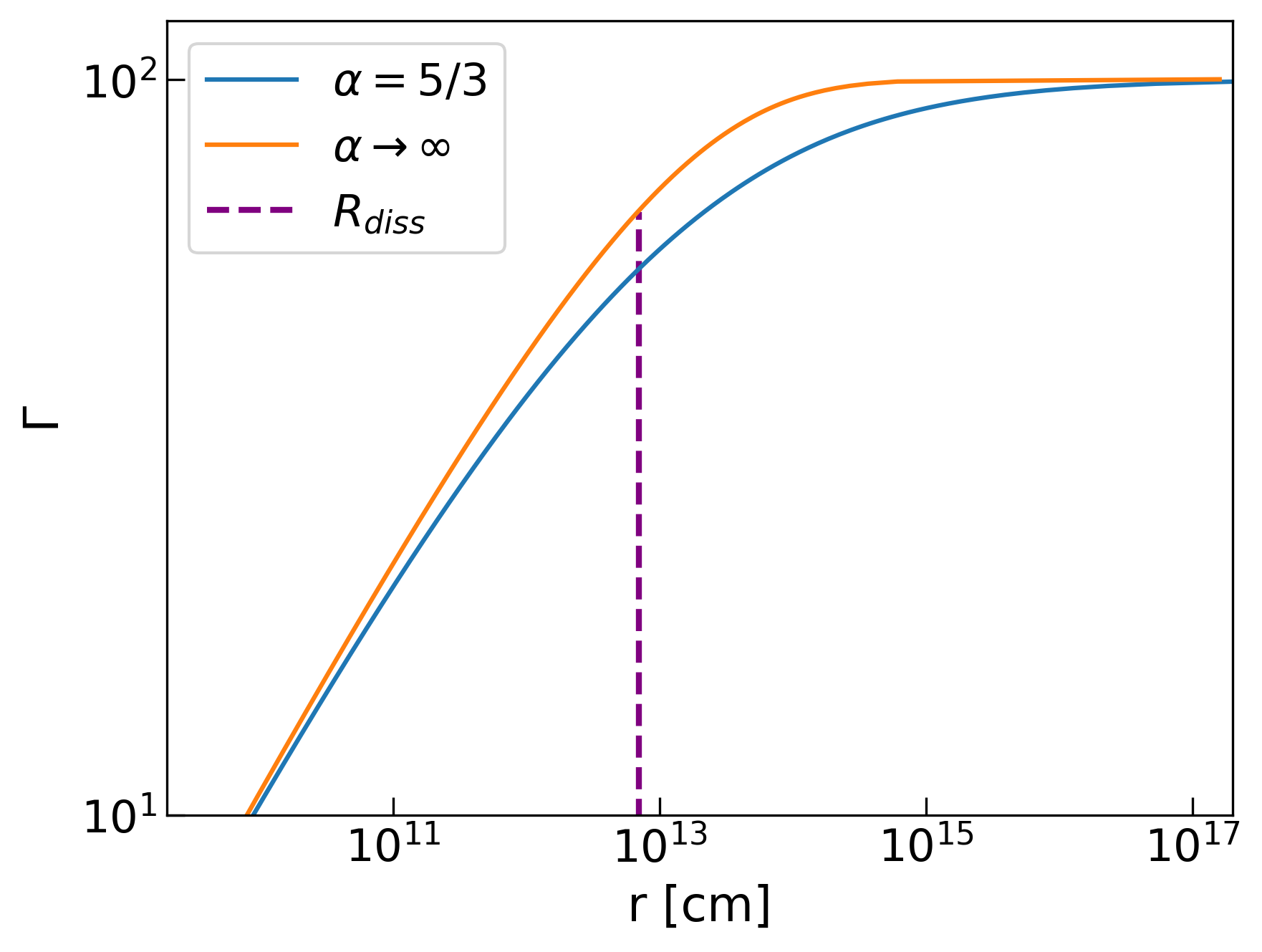}
    \caption{Jet acceleration profile for 2 different values of $\alpha$ (see legend), for $\Gamma_{\infty}=100$ and $R_0= 7\times 10^8$ cm. The dashed line corresponds to the dissipation radius. Note that the two profiles converge for small and large values of r ($r\ll R_{\text{diss}}$ and $r\rightarrow\infty$, respectively), while they differ the most close to the dissipation radius.}
    \label{fig:Gammawitha}
\end{figure}

\begin{figure}
	\includegraphics[width=\columnwidth]{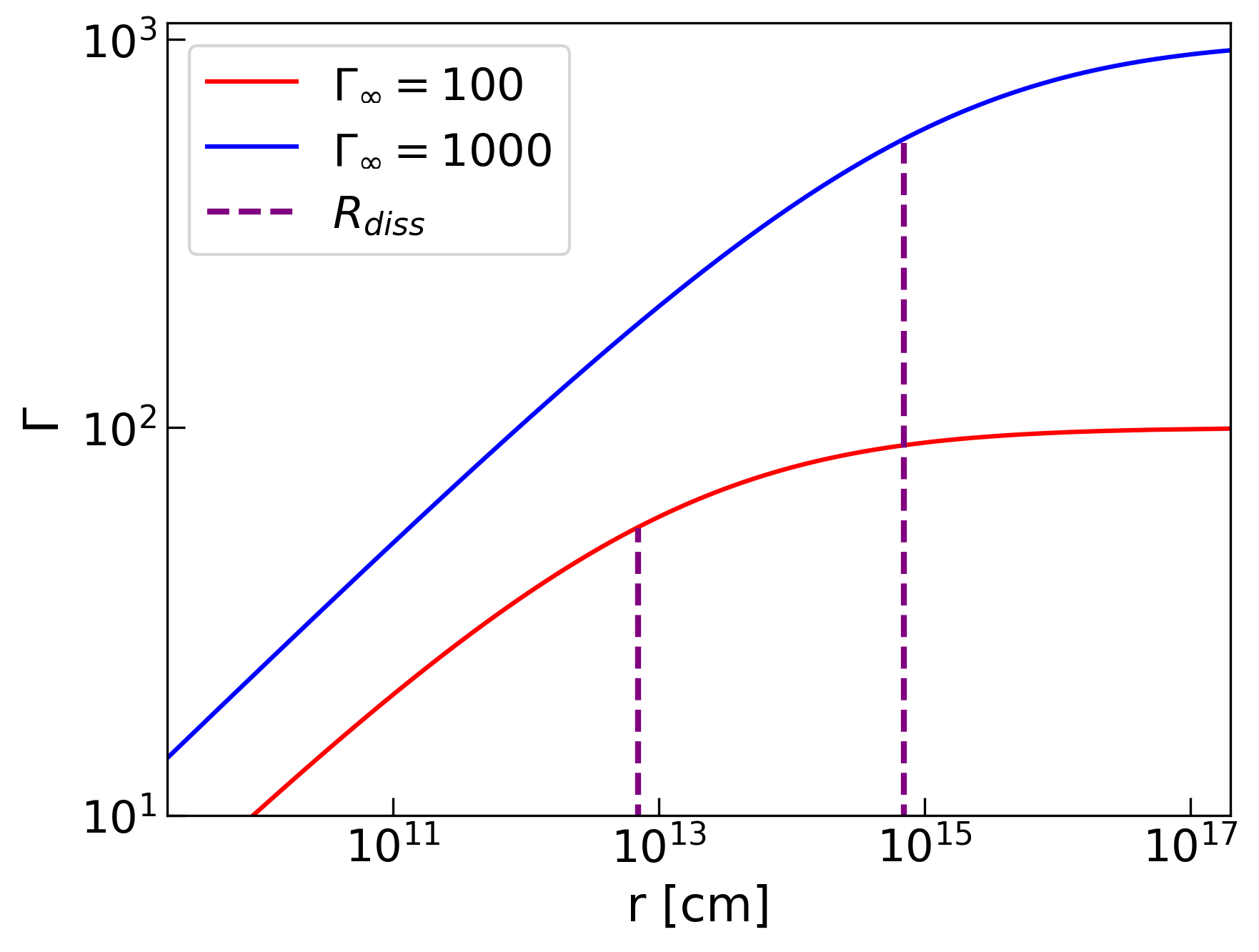}
    \caption{Jet acceleration profile for 2 different values of $\Gamma_{\infty}$ (see legend), for $\alpha=5/3$ and $R_0= 7\times 10^8$ cm. The dashed lines correspond to the dissipation radii for each acceleration profile. Note that the two profiles are self-similar, since their only difference is on the asymptotic Lorentz factor, and in dimensionless units the two profiles are the same.}
    \label{fig:GammawithGinf}
\end{figure}

\begin{figure}
	\includegraphics[width=\columnwidth]{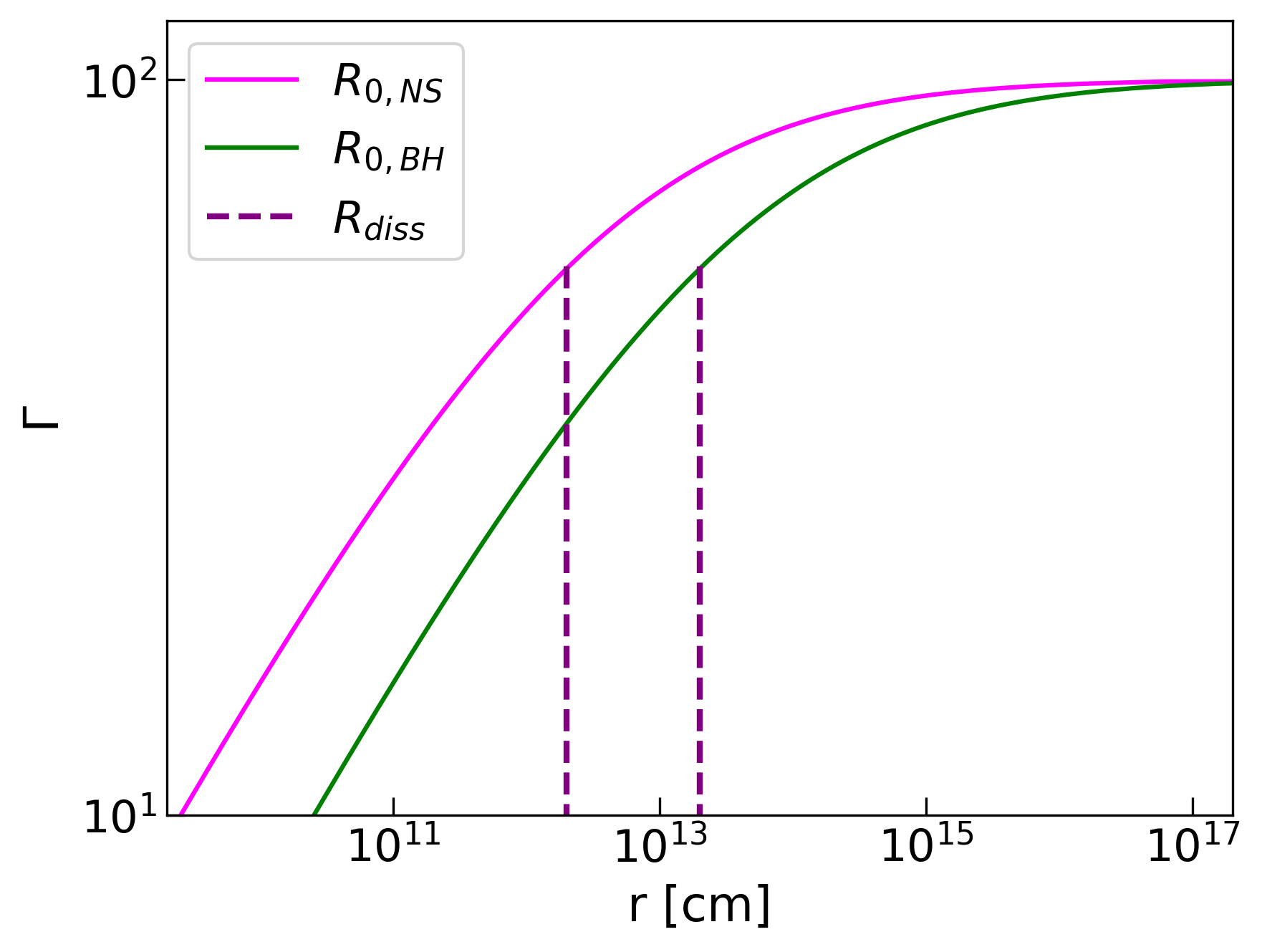}
    \caption{Jet acceleration profile for 2 different values of $R_0$, for $\alpha=5/3$ and $\Gamma_{\infty}=100$. The chosen values of $R_0$ correspond to characteristic values for a neutron star and a black hole ($\sim2\times 10^8$~ cm and $2\times 10^9$~cm, respectively). The dashed lines correspond to the dissipation radii for each acceleration profile.}
    \label{fig:GammawithR0}
\end{figure}

The three parameters $(\Gamma_{\infty}, R_0, \alpha)$ fully determine the behavior of the jet's acceleration profile. Comparing the D02 model (single stripe, $\alpha \rightarrow \infty$) with the multiple-stripe model (small $\alpha$), we see that for both cases the dissipation peaks around $R_{\text{diss}}$; for the former, the dissipation drops fast for $r>R_{\text{diss}}$ where the jet reaches a Lorentz factor close to the asymptotic one, $\Gamma\approx \Gamma_{\infty}$. On the other hand, a model with a small value of $\alpha \sim 1-3$ has a more gradual dissipation and the asymptotic velocity is reached at $r\gg R_{\text{diss}}$ (see Fig. \ref{fig:Gammawitha}). 

By changing the other two parameters of the model ($\Gamma_{\infty}$ and $R_0$), we see that they both affect the distance scale over which the jet accelerates. Larger values of $\Gamma_{\infty}$ and $R_0$ increase the distance from the central engine where dissipation peaks, with the former having a more significant contribution due to the $R_{\text{diss}}\propto R_0\Gamma_{\infty}^2$ dependence of the dissipation radius (see Figs. \ref{fig:GammawithGinf} and \ref{fig:GammawithR0}).

\section{Photospheric emission in the striped wind model}

In the striped jet model, the dissipation of magnetic energy proceeds continuously on a large range of distances with the dissipation rate peaking at $R_{\rm diss}$. Energy dissipated at the optically thick parts of the flow (much below the photospehric radius $R_{\rm ph}$) is effectively thermalized. When $R_{\rm diss}\ll R_{\rm ph}$, most of the dissipated energy is subject to adiabatic losses and the radiative luminosity from the flow is very weak. When dissipation proceeds at $R_{\rm ph}$, spectral distortions are introduced by electron scattering close to the photosphere leading to photospheric emission reminiscent of the observed GRB emission (\citealp{Giannios2006}). The luminosity of the photopsheric component is maximal when $R_{\rm diss}\sim R_{\rm ph}$. On the other hand, when $R_{\rm diss}\gg R_{\rm ph}$, most of the dissipation takes place on the optically thin segments of the flow and the emission is likely to be of synchrotron origin (\citealp{GianniosSpruit2005}). Here, we proceed to calculate the luminosity of the photopsheric and of the optically thin components as well as the spectral peak expected by the photosphere of the flow as a function of the model parameters.

\subsection{Efficiencies}

We calculate the Thomson optical depth following \cite{Abramowicz1991}. The optical depth $d\tau$ in a radial, ultra-relativistic ($\Gamma \gg 1$) flow is given by
\begin{equation}
    \label{eq:dtau}
    d\tau = \frac{n'\sigma_Tdr}{2\Gamma},
\end{equation}
where $n'=\frac{L_{\text{iso}}}{4\pi m_p c^3\Gamma_{\infty}\Gamma r^2}$ is the comoving number density and $L_{\text{iso}}$ is the isotropic luminosity of the burst. Integrating equation (\ref{eq:dtau}) from $r$ to $\infty$ and changing the variables $(\Gamma, r)\rightarrow(\chi, \zeta)$, we obtain
\begin{equation}
    \tau_k(\chi)=\frac{\kappa L_{\text{iso}}}{24\pi c^3\Gamma_{\infty}^5 R_0}\int_{\chi}^{1}\frac{d\chi'}{\zeta_{k}^2(\chi')(1-\chi')^k},
    \label{eq:opticaldepth}
\end{equation}
where $\kappa=\sigma_T/m_{\text{p}}$ is the opacity of the flow. Equation (\ref{eq:opticaldepth}) depends on the value of $\alpha$ and for a given set of the model's parameters $(L_{\text{iso}}, \Gamma_{\infty}, R_0)$ it can be solved to determine the photospheric radius and velocity. More specifically, the photospheric velocity $\chi_{\text{ph}}$ is found by solving the equation $\tau_k(\chi_{\text{ph}})=1$, and the photospheric radius will be simply given by $\zeta_{\text{ph}}\equiv\zeta_k(\chi_{\text{ph}})$. In the two limiting cases of equation (\ref{eq:zetachi}) we have:
\begin{equation}
\zeta_{\text{ph}}=
    \begin{cases}
       \chi_{\text{ph}}^3/3 &\quad\quad \chi_{\text{ph}}\ll1\\
      (1-\chi_{\text{ph}})^{1-k}/k-1 & \quad\quad \chi_{\text{ph}}\simeq 1,\\
    \end{cases}   
\label{eq:zphlimits}
\end{equation}
so by combining equations (\ref{eq:opticaldepth}) and (\ref{eq:zphlimits}) we obtain: 
\begin{equation}
R_{\text{ph}}=
    \begin{cases}
    \left(\frac{3\kappa}{40\pi c^3}\right)^{3/5}\frac{L_{\text{iso}}^{3/5}R_0^{2/5}}{\Gamma_{\infty}} & \quad\quad \text{$R_{\text{ph}}\ll R_{\text{diss}}$}\\
    \frac{\kappa L_{\text{iso}}}{8\pi c^3\Gamma_{\infty}^3}  &\quad\quad\text{$R_{\text{ph}}\gg R_{\text{diss}}$}.\\
    \end{cases}     
\label{eq:rphot}
\end{equation}

The ratio $R_{\text{diss}}/R_{\text{ph}}$ is an important parameter for the photospheric emission.  
From equation (\ref{eq:rdiss}) and equation (\ref{eq:rphot}) we can calculate the ratio of the two characteristic radii $R_{\text{diss}}/R_{\text{ph}}$ in the two regimes:
\begin{equation}
\frac{R_{\text{diss}}}{R_{\text{ph}}}=
    \begin{cases}
     \left(\frac{40 \pi c^3 R_0}{3\kappa} \right)^{3/5} \left(\frac{\Gamma_{\infty}^5}{L_{\text{iso} }}\right)^{3/5}  &\quad\quad\text{$R_{\text{ph}}\ll R_{\text{diss}}$}\\
      \frac{8 \pi c^3 R_0}{\kappa} \frac{\Gamma_{\infty}^5}{L_{\text{iso}}} & \quad\quad \text{$R_{\text{ph}}\gg R_{\text{diss}}$.}\\
    \end{cases}   
\label{eq:RdissRph}
\end{equation}

For constant $R_0$, we have the ratio of the two radii to be proportional to $(\Gamma_{\infty}^5/L_{\text{iso}})^{3/5}$ and $\Gamma_{\infty}^5/L_{\text{iso}}$, respectively, for the two different limits. Also, the transition from the one regime to the other occurs at $R_{0,\text{cr}}=1.3\times10^9\frac{L_{\text{iso},52}}{\Gamma_{\infty,2}^5}\:$ cm, the radius where the two brackets of equation (\ref{eq:RdissRph}) become comparable and close to 1. Here and for the rest of this work we are using the notation $A_x=A/10^x$.
Based on the critical value of $R_0$ and the model parameters ($L_{\text{iso}}, \Gamma_{\infty}, R_0$) we can determine in what regime we are. 

At every distance from the central engine, approximately half of the dissipated magnetic energy goes directly to the bulk acceleration of the flow. Below and above the photosphere, the other half thermalizes and has a non-thermal signature, respectively. To calculate the photospheric efficiency, i.e., the percentage of the total energy in the photosphere that is radiated away, we must also include the adiabatic loses from the thermal energy, which we discuss below. 

For a blackbody, the luminosity and pressure scale as $L(r)\propto r^2 \Gamma^2 T^4$ and $\pi=Nk_BT+\alpha T^4$, respectively. For an adiabatic expansion of a radiation-dominated flow, we have $\alpha T^4 \gg Nk_B T$, so the pressure will scale as $\pi \propto T^4$ and $\pi \propto \rho^{4/3}$. Combining the last two equations we obtain $T\propto \rho^{1/3}$. Assuming a cold flow and integrating the ideal MHD conservation laws we find $\rho=\frac{L}{\Gamma_{\infty}c^3r^2 \Gamma}\propto r^{-2}\Gamma^{-1}$ 
Combining the expressions above, we obtain $T\propto r^{-2/3}\Gamma^{-1/3}$ and $L\propto r^{-2/3}\Gamma^{2/3}$. If $L(r)$ is the thermal luminosity of the jet at a distance $r<R_{\text{ph}}$, then only a fraction $f$ of that luminosity will remain in thermal form in the photosphere; the rest will be lost due to the adiabatic expansion of the jet. We can write this fraction as $f(r)=\frac{dL_{\text{ph}}(r)}{dL(r)}$, or in terms of the dimensionless variables $\zeta$ and $\chi$, $dL_{\text{ph}}(\zeta,\zeta+d\zeta)=f(\chi)L\frac{d\chi}{d\zeta}d\zeta$, where $f(\chi)$ is defined as 
\begin{equation}
    \label{eq:adiabatic}
    f(\chi)=\left(\frac{\zeta_k(\chi)}{\zeta_{\text{ph}}} \right)^{2/3}\left( \frac{\chi}{\chi_{\text{ph}}}\right)^{-2/3}, 
\end{equation}
 and is valid for $\zeta<\zeta_{\text{ph}}$. We assume that $f(\chi)=1$ for $\zeta>\zeta_{\text{ph}}$, that is, there are no adiabatic losses above the photosphere. 

The dissipation profile is given by
\begin{equation*}
   \frac{dL_{\text{diss}}}{dr}=\frac{d}{dr}(\Gamma \dot{M}c^2)=\frac{L}{\Gamma_{\infty}}\frac{d\Gamma}{dr},
\end{equation*}
or using equation (\ref{eq:xzeq}) we can write
\begin{equation}
    \frac{dL_{\text{diss}}}{d\zeta}=L\frac{d\chi}{d\zeta}=L\frac{(1-\chi)^k}{\chi^2}
\end{equation}
in terms of $\zeta$ and $\chi$. To find the total dissipated energy up to the photospheric radius, we include the adiabatic loses through equation (\ref{eq:adiabatic}) and the fact that approximately 50 per cent of the energy is used to accelerate the flow to obtain
\begin{equation}
    \label{eq:Lph}
    L_{\text{ph}}=\frac{L_{\text{iso}}}{2}\int_0^{x_{\text{ph}}}f(\chi)d\chi,
\end{equation}
for the thermal energy below the photosphere, and
\begin{equation}
    \label{eq:Lnt}
    L_{\text{nt}}=\frac{L_{\text{iso}}}{2}(1-\chi_{\text{ph}}),
\end{equation}
for an upper limit on the non-thermal dissipated energy above the photosphere. The photospheric and non-thermal efficiencies will be simply $\epsilon_{\text{ph}} = L_{\text{ph}}/L_{\text{iso}}$ and $\epsilon_{\text{nt}} = L_{\text{nt}}/L_{\text{iso}}$, respectively. We see that the photospheric and non-thermal efficiency of the GRB will depend on the parameters $(L, \Gamma_{\infty}, R_0, \alpha)$ indirectly through the value of $\chi_{\text{ph}}$.

\begin{figure}
	\includegraphics[width=\columnwidth]{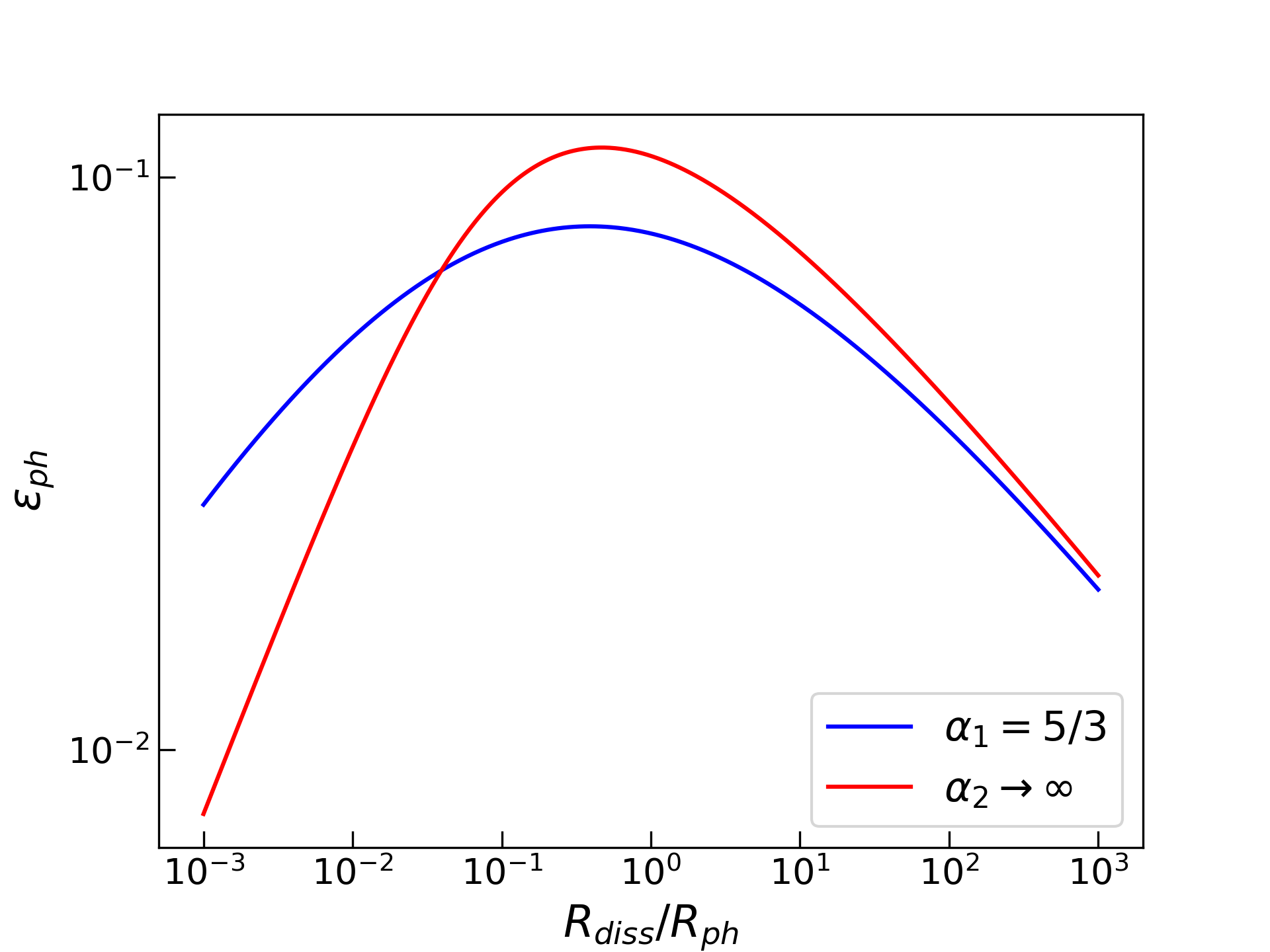}
    \caption{Photospheric efficiency as a function of $R_{\text{diss}}/R_{\text{ph}}$ for $\alpha_1=5/3$ (blue) and $\alpha_2\rightarrow\infty$ (red). Note that the maximum efficiency depends on the value of $\alpha$ but it always happens when the two radii are comparable. Furthermore, the smaller the value of $\alpha$ is, the less sensitive the dependence of the photospheric efficiency on $R_{\text{diss}}/R_{\text{ph}}$.} 
    \label{fig:thermaleff}
\end{figure}

In a similar way to equation (\ref{eq:RdissRph}), we obtain from equation (\ref{eq:Lph}) in the two limits:
\begin{equation}
\epsilon_{\text{ph}}=
    \begin{cases}
    \frac{3}{14}\left(
    \frac{R_{\text{diss}}}{R_{\text{ph}}}\right)^{-1/3} &\quad\quad \text{$R_{\text{ph}}\ll R_{\text{diss}}$}\\
    \frac{3}{14}\left(\frac{R_{\text{diss}}}{R_{\text{ph}}} \right)^{2/3}   &\quad\quad \text{$R_{\text{ph}}\gg R_{\text{diss}},$}\\
    \end{cases}       
\label{eq:eph}
\end{equation}
where we have used equation (\ref{eq:zphlimits}) for the two limits of the efficiency.

The first limit of equation (\ref{eq:eph}) is valid for any value of the stripe parameter $\alpha$, because for $R_{\text{diss}}/R_{\text{ph}}\gg 1$ all solutions converge to the D02 limit ($\Gamma \propto r^{1/3}$), where the acceleration profile is independent of $\alpha$ and the contribution of the smaller stripes dominates. 
On the other hand, the second limit is accurate only for $\alpha \gg 1$. In general, for $R_{\text{diss}}/R_{\text{ph}}\ll 1$ the photospheric efficiency is set by the asympotic dissipation and $\epsilon_{\text{ph}}\propto (R_{\text{diss}}/R_{\text{ph}})^{b(\alpha)}$, where $b\rightarrow 2/3$ for $\alpha \rightarrow \infty$, in agreement with the Drenkhahn Limit, and $b\rightarrow 1/2$ for $\alpha \rightarrow 5/3$. This occurs because in that regime we have multiple stripe lengths that contribute to the residual dissipation, with a distribution that depends on $\alpha$. Consequently, the thermal efficiency in the multiple-stripe model will depend only on $\alpha$ and $R_{\text{diss}}/R_{\text{ph}}$.
In Fig. \ref{fig:thermaleff} we present this dependence for two different values of the stripe parameter. 

\begin{figure}
	\includegraphics[width=\columnwidth]{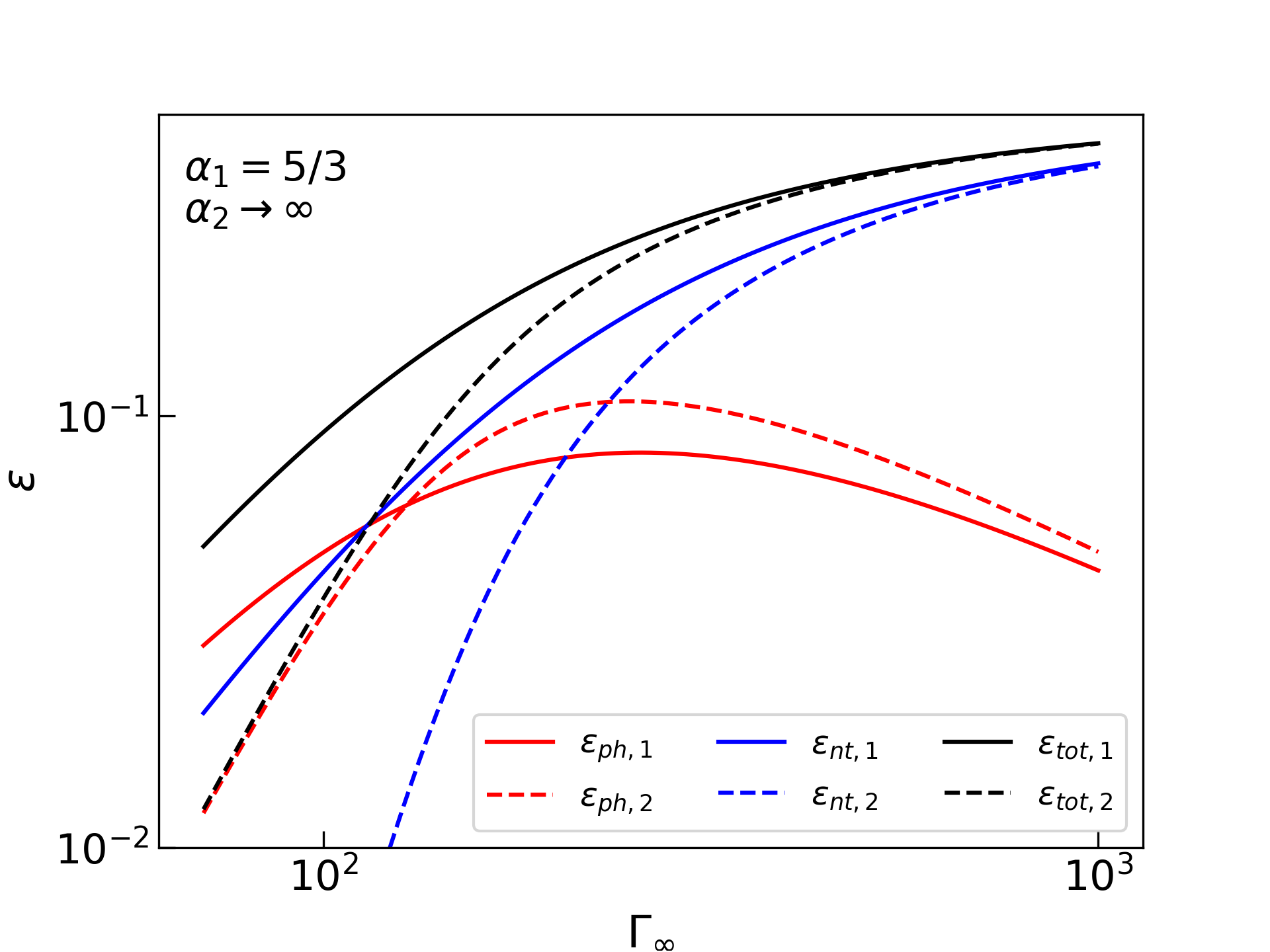}
    \caption{Photospheric (red), non-thermal (blue) and total (black) efficiencies as a function of the asymptotic Lorentz factor for $R_0=10^8 \:\text{cm}, \: L_{\text{iso}}=10^{53}\:$erg/s. The continuous lines correspond to $\alpha_1=5/3$, and the dashed lines to $\alpha_2\rightarrow\infty$.}
    \label{fig:theoreticaleff}
\end{figure}

In Fig. \ref{fig:theoreticaleff}, we also plot the photospheric (red) and upper limit of non-thermal (blue) efficiencies along with the total efficiency $\epsilon_{\text{ph}}+\epsilon_{\text{nt}}$ (black) as a function of $\Gamma_{\infty}$ for constant values of $R_0,L_{\text{iso}}$ and two different values of the stripe parameter $\alpha$. The values of $\epsilon_{\text{ph}}, \epsilon_{\text{nt}}$ are consistent with \cite{Peer2017}, where their upper limits have been estimated to be 30 and 50 per cent, respectively.
We notice that the photospheric efficiency has a behavior similar to the one of Fig. \ref{fig:thermaleff} for the two values of $\alpha$, where it reaches a maximum value of 8-10 per cent at $\Gamma_{\infty}\approx 250$ or  $R_{\text{diss}}/R_{\text{ph}}\sim 1$ as we can see from equation (\ref{eq:RdissRph}). On the other hand, the non-thermal efficiency increases monotonically with $\Gamma_{\infty}$, and in the limit $\Gamma_{\infty}\rightarrow\infty, R_{\text{ph}}\ll R_{\text{diss}}$ it reaches 50 per cent, which corresponds to the total available amount of energy given.





\subsection{Peak Energy}
\label{section:peakenergy}

The magnetic energy in the flow starts dissipating at small distances $r\ll R_{\text{diss}}$ ($\zeta\approx \chi^3/3 \ll 1$) and a significant fraction of it converts to internal flow energy by the dissipation radius $R_{\text{diss}}$ ($\zeta_{\text{diss}}=1/3$). The remaining magnetic energy dissipates at large distances where $r\gg R_{\text{diss}}$ and $\Gamma \approx \Gamma_{\infty}$ ($\zeta\gg 1$ and $\chi\approx 1$). 

Depending on the model parameters, the photosphere could be above or below that dissipation radius, and the dissipation either stops at optically thick regions ($\tau\gg 1$), or it transitions to optically thin regions ($\tau\ll1$) in a continuous way. In both cases, the photospheric luminosity will be given by equation (\ref{eq:Lph}), where once again we assume that we have no adiabatic loses for the radiation above the photospheric radius $R_{\text{ph}}$. Since our flow is radiation-dominated, the thermal energy density will be given by $U'_{\text{ph}}=\frac{4\sigma_{\text{SB}}}{c}T_{\text{ph}}^{'4}$, and the thermal luminosity scales with the distance as $  L_{\text{ph}}=\frac{16\pi}{3}R_{\text{ph}}^2 \Gamma_{\text{ph}}^2 c U'_{\text{ph}}$. Combining the two equations, we obtain the comoving temperature of the flow at the photosphere:
\begin{equation}
    \label{eq:kTph}
    k_{\text{B}} T'_{\text{ph}}=k_B \left( \frac{3L_{\text{ph}}}{64\pi\sigma_{\text{SB}}R_{\text{ph}}^2\Gamma_{\text{ph}}^2}\right)^{1/4}
\end{equation}
and the peak energy at the rest frame will be
\begin{equation}
 E_{\text{p,0}}=(1+z)E_{\text{p,obs}}(R_{\text{ph}})=\Gamma_{\text{ph}}3k_{\text{B}}T'_{\text{ph}},    
\end{equation}
or in terms of dimensionless quantities at the photospheric radius:
\begin{equation}
    \label{eq:EpT}
    E_{\text{p,0}}=\left(\frac{3^3k_{\text{B}}^4}{2^5\sigma_{\text{SB}}} \right)^{1/4} \left(\frac{L_{\text{iso}}}{R_0^2\Gamma_{\infty}^2}\right)^{1/4}\left(\frac{\chi_{\text{ph}}}{\zeta_{\text{ph}}}\right)^{1/2}\left(\int_{0}^{\chi_{\text{ph}}}f(\chi)d\chi\right)^{1/4}.
\end{equation}
Equation (\ref{eq:EpT}) gives the $Ef(E)$ peak of the outflow at the photospheric radius. In the two regimes considered above we find 
\begin{equation}
\label{eq:EpTlimits}
E_{\text{p,0}}\propto
    \begin{cases}
    L_{\text{iso}}^{1/10}\Gamma_{\infty}^{1/4}R_0^{-7/20} &\quad\quad \text{$R_{\text{ph}}\ll R_{\text{diss}}$}\\
    L_{\text{iso}}^{-5/12}\Gamma_{\infty}^{17/6}R_0^{1/6}  &\quad\quad \text{$R_{\text{ph}}\gg R_{\text{diss}}$}.\\
    \end{cases}       
\end{equation}

We should note here that although we are referring to the peak energy and the photospheric efficiency as `thermal', the spectrum is modified due to the inverse Compton scattering of the electrons, and it is not a blackbody. If we want to further improve our estimate for the peak energy we can add the effects of particle distribution in our radiative transfer calculations. This can happen in several ways, including energy injection from a non-thermal distribution of electrons, or distributed heating of particles due to magnetic reconnection and subsequent turbulence in the flow. In \cite{Giannios2007} where the latter is considered, a high energy tail appears in the spectrum as a result of the inverse Compton scattering of the hot electrons at the $\tau\sim 1$ region of the jet with the thermal photon background. The modified thermal spectrum of such process represents the GRB prompt data more accurately and gives the following expression for the peak energy \citep{Giannios2007}
\begin{equation}
    E_{\text{p}}=36 L_{\text{iso},52}^{-0.33}\Gamma_{\infty,3}^{2.4}R_{0,8}^{0.11}(\cosh q)^{-0.55} \text{MeV},
    \label{eq:Epeak}
\end{equation}
where $q=6 \ln \left[3.84 \left(\frac{\Gamma_{\infty,3}^5R_{0,8}}{L_{\text{iso},52}}\right)^{2/15}\right]$ depends sensitively on $R_{\text{diss}}/R_{\text{ph}}$ as we can see from equation (\ref{eq:RdissRph}). Similarly to equation (\ref{eq:EpTlimits}), the behavior of the peak energy for large and small values of $R_{\text{diss}}/R_{\text{ph}}$ is, respectively:
\begin{equation}
E_{\text{p}}\propto
    \begin{cases}
    L_{\text{iso}}^{0.11}\Gamma_{\infty}^{0.2}R_0^{-0.33} &\quad\quad \text{$R_{\text{ph}}\ll R_{\text{diss}}$}\\
    L_{\text{iso}}^{-0.77}\Gamma_{\infty}^{4.6}R_0^{0.55}  &\quad\quad \text{$R_{\text{ph}}\gg R_{\text{diss}}$.}\\
    \end{cases}    
\label{eq:Epeaklimits}
\end{equation}

Comparing equations (\ref{eq:EpTlimits}) and (\ref{eq:Epeaklimits}), we see that there is a good agreement between the power-law dependencies of the three parameters ($L_{\text{iso}},\Gamma_{\infty},R_0$) for $R_{\text{ph}}\ll R_{\text{diss}}$ but in the limit of $R_{\text{ph}}\gg R_{\text{diss}}$ the power-laws differ significantly. For the rest of this work we will use equation (\ref{eq:Epeak}) for the calculation of the peak energy.  

The dependence of the peak energy $E_{\text{p}}$ with $R_0$ is not monotonic. As can be seen from equation (\ref{eq:Epeaklimits}), for small values of $R_0$ and $R_{\text{diss}}/R_{\text{ph}}$, $E_{\text{p}}$ increases as $E_{\text{p}}\propto R_{0}^{0.55}$, while for larger values it drops as $E_{\text{p}}\propto R_{0}^{-0.33}$. This implies the existence of a maximum peak energy (in terms of $R_0$, and for $L_{\text{iso}}$ and  $\Gamma_{\infty}$ constant), $E_{\text{p,max}}=12L_{\text{iso},52}^{-0.22}\Gamma_{\infty,3}^{1.85}$ MeV at $R_{\text{0,max}}=5.7\times 10^8\frac{L_{\text{iso},52}}{\Gamma_{\infty,2}^5}$ cm. $R_{0,\text{max}}$ has the same physical meaning as $R_{0,\text{cr}}$ and corresponds to the value of $R_0$ where the photospheric and dissipation radii are comparable. As a result, for $R_0<R_{0,\text{max}}$ we have $R_{\text{ph}}>R_{\text{diss}}$ and the emission is sub-photospheric, while for $R_0>R_{0,\text{max}}$ we have the opposite. 



\section{Application to GRB data}

We can now apply the theory discussed in the previous sections to two different sets of observational constraints. Each data set is associated with a particular sample of GRBs: the "Golden Sample" (G-Sample) and the "Silver Sample" (S-Sample).

The G-Sample consists of a small number of bursts (4 bursts) with well-followed GRB afterglows, and its parameters have been taken from \cite{Laskar2013,Laskar2018c,Laskar2018,Laskar2019} (see Table \ref{tab:GStable}). As a result, both the Lorentz factor ($\Gamma_{\text{dec}}\equiv \Gamma(R_{\text{dec}})$) and the magnetization ($\sigma_{\text{dec}}\equiv \sigma(R_{\text{dec}})$) of the ejecta at the deceleration radius $R_{\text{dec}}$ could be determined, where  
\begin{equation}
    R_{\text{dec}}=\left(\frac{(3-s)E_{\text{iso}}}{4\pi n_0 m_{\text{p}} c^2 \Gamma_{\text{dec}}^2} \right)^{\frac{1}{3-s}},
    \label{eq:rdec}
\end{equation}
is defined as the radius where the jet accumulates an external medium mass equal to $\sim 1/\Gamma$ of its own mass (\citealp{ReesMeszaros1992,Dermer1999}). The combination of these two quantities ($\Gamma_{\text{dec}}, \sigma_{\text{dec}}$) can provide valuable information about the jet, such as the stripe distribution of the magnetic field $\alpha$. For equation (\ref{eq:rdec}), $E_{\text{iso}}$ is the isotropic kinetic energy of the jet and $s$ is the parameter that determines the density profile of the external medium, $n=n_0/r^s$. The values $s=0$ and $s=2$ correspond to a constant density medium and wind density profile, respectively. For $s=0$, $n_0$ is the number density, and for $s=2$, $n_0=5\times10^{11}A_{*} \text{ g/cm}$, where $A_{*}=(\dot{M}/10^{-5}M_{\odot}\text{yr}^{-1})(v_w/1000\,\text{km}\,\text{s}^{-1})^{-1}$ (\citealp{ChevalierLi2000}). 

The S-Sample consists of a larger number of bursts (67) and it is taken from \cite{Ghirlanda2018}. Each GRB on the S-Sample has an estimate of $\Gamma_{\text{dec}}$ (calculated in \citealp{Ghirlanda2018}), but lacks an estimate of $\sigma_{\text{dec}}$. For this reason we cannot estimate a value of $\alpha$ for the S-Sample. Instead, we can assume a value for $\alpha$ based on the G-Sample, and proceed to calculate the photospheric peak energy and efficiencies for these bursts. As shown above, these quantities do not sensitively depend on the stripe parameter $\alpha$, so the value of $\alpha$ from the G-Sample simply serves as a guideline.

For our calculations of the photospheric quantities $R_{\text{ph}}$ and $\Gamma_{\text{ph}}$, we can approximate $\Gamma_{\infty}\approx \Gamma_{\text{dec}}$, where $\Gamma_{\text{dec}}$ is the Lorentz factor at the deceleration radius.
We can justify that approximation either from equation (\ref{eq:gammasigma}), where $\sigma_{\text{dec}}\ll 1$ as we can see for our G-Sample in Table \ref{tab:GStable}, or from comparing equations (\ref{eq:rdiss}) and (\ref{eq:rdec}), where we find that $R_{\text{diss}} \approx 10^{14} \text{cm} \, R_{0,8}\Gamma_{\infty,3}^2$ $\ll R_{\text{dec}} \approx 10^{17} \text{cm}$ so that most of the acceleration takes place well below the deceleration radius. 


\subsection{Golden Sample: Known values of $\Gamma_{\text{dec}}$ and $\sigma_{\text{dec}}$}
\label{section:GoldenSample}

We start with the G-Sample, a subset of GRB that have the following quantities obtained by detailed afterglow modeling: $(z, E_{\text{iso}}, T_{90}, \Gamma_{\text{dec}}, E_{p,\text{obs}}, \sigma_{\text{dec}}, n_0, s)$.  The data for this sample are collected from the detailed analysis of \cite{Laskar2013,Laskar2018c,Laskar2018,Laskar2019}. Here, $T_{90}$ is the duration of the burst, taken as the time interval in which the integrated counts increase from 5 to 95 per cent of the total counts. 
The difference of the G-Sample that separates it from the S-Sample is that we have good quality afterglow observations that allow for the identification of the reverse shock and the estimation of the afterglow microphysical parameters $\varepsilon_{\text{B,f}}$, $\varepsilon_{\text{B,r}}$, which are the fractions of shocked energy in magnetic fields in the forward and the reverse shock, respectively. With these parameters, the magnetization of the ejecta at the deceleration radius $\sigma_{\text{dec}}$ can be determined. Via afterglow modeling (see Table \ref{tab:GStable}),
 the value of $\Gamma_{\text{dec}}$ has also been determined. 
 
Using the G-Sample data, we can constrain our model parameters and connect the afterglow phase of the burst with the jet properties responsible for the prompt emission mechanism. In order to do this, we need to need to assume a value of $R_0$.
Admittedly, as we move closer to the GRB central engine it becomes more difficult to infer its properties via observations. Therefore, the nature of the central engine and consequently the value of $R_0$ is unknown. As described in Section \ref{section:R0}, we consider two possible central engines (NS or BH) with a range of values of $R_0$ for each. We use $R_0$ in this section as a free parameter. This will allow us to estimate the stripe parameter $\alpha$ for these bursts as a function of $R_0$ (the central engine) as described below.
 

\begin{table*}
    \centering
    \renewcommand{\arraystretch}{1.2}
    \begin{tabular}{cccccccccc}
        \hline
        &z&$\log E_{\text{iso}}$&$T_{90}$(s)&$\Gamma_{\text{dec}}$&$E_{\text{p,obs}}$ (keV)&$\sigma_{\text{dec}}$&$n_0$ &$A_*$&Refs.\\
        \hline
        \hline
        \\[-1em]
        $\text{GRB 181201A}$ &$0.45$&$53.41^{+0.13}_{-0.16}$&$19.2\pm 2.25^{\dag}$&$103^{+10}_{-8}$&$152\pm 6$&$(2.6\pm 1.3^{\dag})\times 10^{-3}$&&$1.9\times 10^{-2}$&$1$\\
        \\[-1em]
        \hline
        \\[-1em]
        $\text{GRB 161219B}$ &$0.1475$&$51.66^{+0.12}_{-0.09}$&$6.94\pm 0.79$&$100 \pm 10^{\dag}$&$61.9\pm 16.5$&$0.06\pm 0.03^{\dag}$&$3\times 10^{-4}$&&$2$\\
        \\[-1em]
        \hline
        \\[-1em]
        $\text{GRB 140304A}$&$5.283$&$54.69^{+0.13\dag}_{-0.12}$&$15.6\pm 1.9 $&$300 \pm 30^{\dag}$&$123\pm 27$&$0.02\pm 0.01^{\dag}$&$1$&&$3$\\
        \\[-1em]
        \hline
        \\[-1em]
        $\text{GRB 130427A}$&$0.34$&$53.3^{+0.13\dag}_{-0.12}$ &$163\pm 19.2^{\dag}$&$130 \pm 13^{\dag}$&$830\pm145^{\dag}$&$0.2\pm 0.1^{\dag}$&&$2.9\times 10^{-3}$&$4$\\
        \\[-1em]
        \hline
        \\[-1em]
    \end{tabular}
    \caption{Parameters of the Golden sample (G-Sample) bursts.
    Calculated parameters were obtained by afterglow modeling performed in the references provided. Errors in some quantities were not reported, for these we estimated them using the fractional error of the reported quantities and marked them with ($^{\dag}$); see Section \ref{section:GoldenSample} for more details. [1]: \protect\cite{Laskar2019}, [2]: \protect\cite{Laskar2018}, [3]: \protect\cite{Laskar2018c}, [4]: \protect\cite{Laskar2013}.}
    \label{tab:GStable}
\end{table*}

We begin by discussing the magnetization $\sigma_{\text{dec}}$. The relative magnetization is defined as $R_B\equiv \sqrt{\frac{\varepsilon_{\text{B,r}}}{\varepsilon_{\text{B,f}}}}$ (e.g., \citealt{ZhangKobayashi2005}). 
Via afterglow modeling, an estimate for $\sigma_{\text{dec}}$ (e.g., \citealt{ZhangKobayashi2005}) can be obtained as follows
\begin{equation}
    \sigma_{\text{dec}} \approx \varepsilon_{\text{B,r}}=R_B^2 \varepsilon_{\text{B,f}},
    \label{eq:sigma}
\end{equation}
where we have assumed a non-relativistic reverse shock.

The value of $\sigma_{\text{dec}}$ is crucial in the calculation of $\alpha$, because  we can calculate the normalized Lorentz factor at $R_{\text{dec}}$ from equation (\ref{eq:gammasigma}):
\begin{equation}
    \label{eq:chidec}
    \chi_{\text{dec}}=\frac{\Gamma_{\text{dec}}}{\Gamma_{\infty}}=\frac{1}{1+\sigma_{\text{dec}}}.
\end{equation}
The dimensionless deceleration radius $\zeta_{\text{dec}}=\frac{R_{\text{dec}}}{3R_{\text{diss}}}$ can be calculated by using equation (\ref{eq:rdec}), the assumed value of $R_0$ and the rest of the model parameters ($s,\Gamma_{\text{dec}},E_{\text{iso}},n_0$). Using those values, and the fact that $\Gamma_{\text{dec}}\approx \Gamma_{\infty}$ (since $\sigma_{\text{dec}}\ll 1$), we have
\begin{equation}
   \label{eq:zetadec}
   \zeta_{\text{dec}}=\frac{1}{3R_0\Gamma_{\infty}^2}\left(\frac{(3-s)E_{\text{iso}}}{4\pi m_{\text{p}} n_0 c^2 \Gamma_{\infty}^2}\right)^{\frac{1}{3-s}}.
\end{equation}
With $\zeta_{\text{dec}}$ and $\chi_{\text{dec}}$ at the deceleration radius, we can replace them back in equation (\ref{eq:zetachi}), and solve an equation of the form $\zeta_k(\chi_{\text{dec}})=\zeta_{\text{dec}}$, where $k$ is the only unknown\footnote{The function $\zeta_k(\chi)$ is continuous and monotonic both in terms of $k$ and $\chi$; because of the former, we can use the main branch of the function in this calculation and ignore the cases where $k={1,2,3}$. The latter ensures that there is only one value of $k$ that satisfies the boundary conditions, that is, only one stripe distribution profile can achieve a specific Lorentz factor for the jet at a specific distance.}; as a reminder $k = (1-3\alpha)/(2-2\alpha)$.  

\begin{figure}
	\includegraphics[width=\columnwidth]{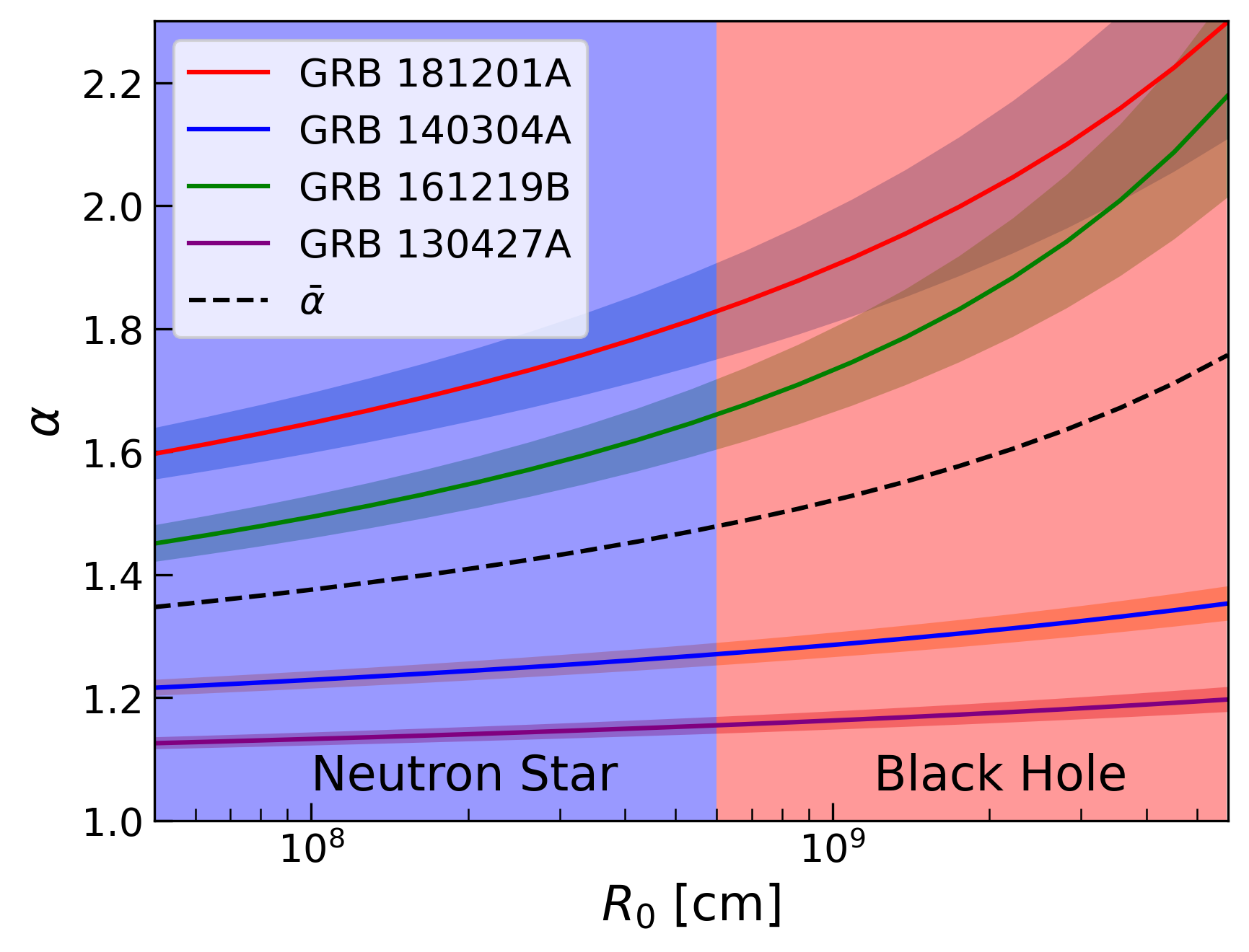}
    \caption{Stripe parameter $\alpha$ as a function of $R_0$ for the GRBs of the Golden Sample. The blue and red areas correspond to the $R_0$ ranges for a NS and BH engine, respectively, as discussed in Section \ref{section:R0}. The shaded areas around the lines correspond to the uncertainty of $\alpha$ given by error propagation.
    Note that the striped jet model favors only small values of $\alpha$ ($<3$). Black hole  models (larger $R_0$) favor slightly larger values of $\alpha$ than those for neutron stars.}
    \label{fig:alphaR0GS}
\end{figure}

From equation (\ref{eq:zetadec}) we see that the calculated value of $\alpha$ will depend on the assumed value of $R_0$. We study the dependence of the stripe parameter $\alpha=\alpha(R_0)$ for a range of $R_0$ from $\text{min}(R_{\text{0,NS}})$ to $\text{max}(R_{\text{0,BH}})$ for each of the G-Sample bursts. The results are presented in Fig. \ref{fig:alphaR0GS}. The uncertainties of $\alpha$ (shaded areas) have been estimated by propagating the equations for $\chi_{\text{dec}}$ and $\zeta_{\text{dec}}$, which depend on the uncertainties of $\sigma_{\text{dec}}$, and $L_{\text{iso}}, \Gamma_{\text{dec}}$, respectively. In this figure, we also plot the function $\bar{\alpha}(R_0)$, which is defined as the average value of $\alpha$ for every value of $R_0$, based on the G-Sample data. This average value will be used later in the study of the S-Sample. From a theoretical point of view, the dependence of $\alpha$ on $R_0$ (or equivalently the dependence of $l_{\text{min}}$, since $R_0=l_{\text{min}}/6\epsilon$) can be translated as an uncertainty of the central engine characteristics (black hole mass/spin and neutron star frequency) as it was discussed in detail on section \ref{section:R0}.

We see that our multiple stripe model favors small values of $\alpha$ ($\alpha<3$) for both types of central engines, which correspond to a more gradual dissipation profile contrary to the single stripe model \citep{Drenkhahn2002}, which is steeper and corresponds to $\alpha\rightarrow \infty$. We also notice that $\alpha$ monotonically increases with $R_0$, so that neutron stars with smaller values of $R_0$ tend to have slightly smaller values of $\alpha$ than black holes.

For the G-Sample we also calculate the theoretical peak energies of the bursts $E_{\text{p}}$ using equation (\ref{eq:Epeak}), and we compare them with the observed values $E_{\text{p,obs}}$ given in Table \ref{tab:GStable}. To calculate the approximate value of $L_{\text{iso}}$ for our sample we are using the first three quantities of the table ($z,E_{\text{iso}},T_{90}$) to find:
\begin{equation}
    L_{\text{iso}}=(1+z)\frac{E_{\text{iso}}}{T_{90}}.
\end{equation}

\begin{figure}
	\includegraphics[width=\columnwidth]{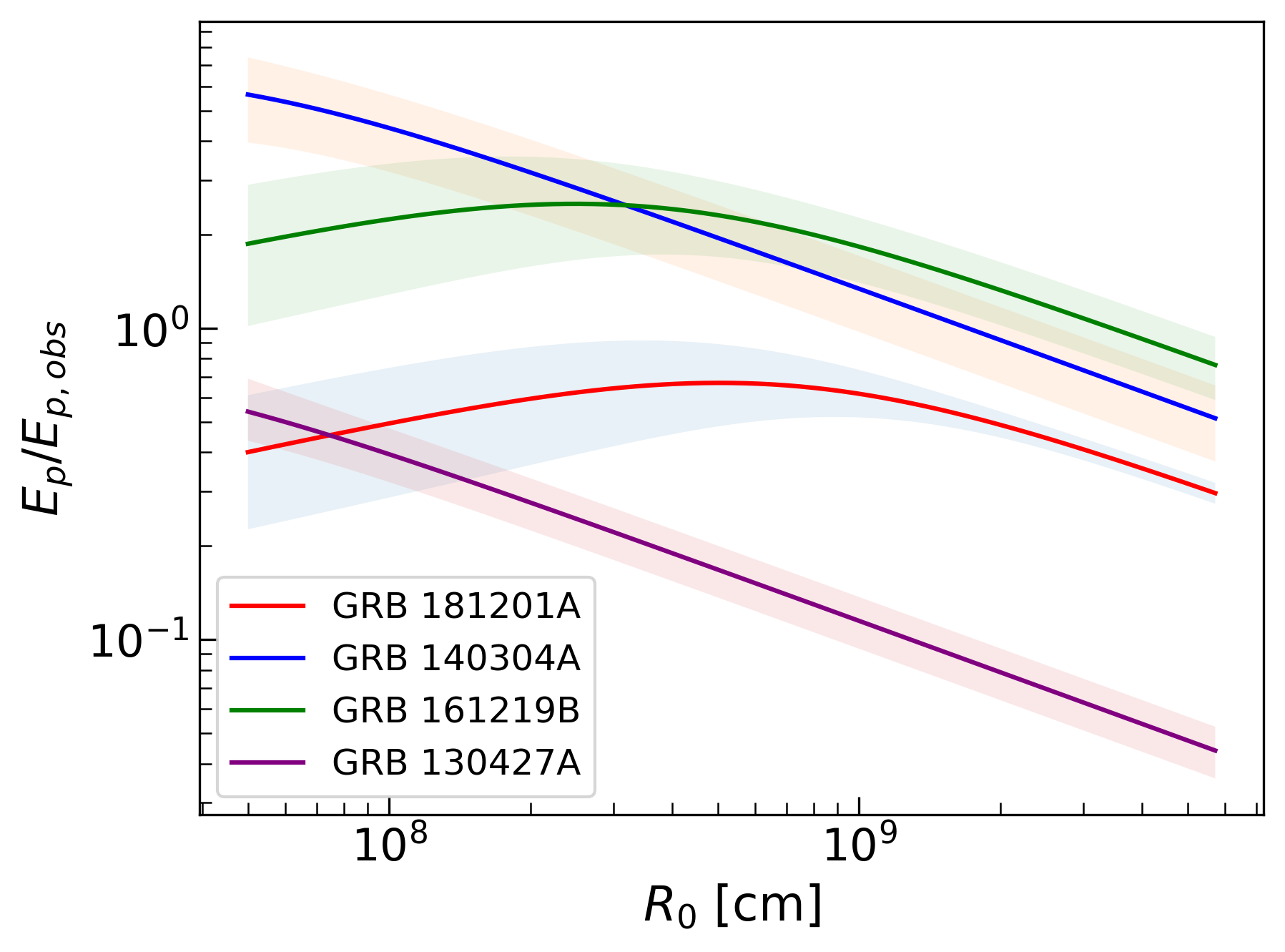}
    \caption{The ratio of the theoretical peak energy from the GRB photosphere $E_{\text{p}}$ to the observed values $E_{\text{p,obs}}$ as a function of $R_0$ for the Golden Sample bursts. The colored areas around the lines correspond to the uncertainty of the ratio given by error propagation. GRBs 181201A and 161219B are close to the maximum theoretical peak energy and the ratio is close to unity for all values of $R_0$, while the other have, for the range of $R_0$ of interest, $R_0>R_{\text{0,max}}$ where the peak energy drops with $R_0$ and $R_{\text{diss}}>R_{\text{ph}}$. In all bursts the theoretical peak energy agrees with the observed one within a factor of 2 over the allowed range of $R_0$.}
    \label{fig:EpeakratioGS}
\end{figure}

Since $R_0$ is a free parameter of the model, we calculate $E_{\text{p}}$ for all the values of $R_0$ in the neutron star and black hole regime. In Fig. \ref{fig:EpeakratioGS} we plot the ratio $E_{\text{p}}/E_{\text{p,obs}}$ as a function of $R_0$ for the bursts of the G-Sample, along with their corresponding uncertainties. We see that for all bursts we estimate the theoretical peak energy within an order of magnitude from the observed value (that is, $E_{\text{p}}/E_{\text{p,obs}} \sim 1$), but their behavior as a function of our chosen range of $R_0$ changes slightly depending on the burst. For GRBs 140304A and 130427A we have $R_0>R_{\text{0,max}}$ which from equation (\ref{eq:Epeaklimits}) corresponds to significant dissipation above the photosphere, and the peak energy drops as we move from a neutron star to a black hole engine. On the other hand, for GRBs 181201A and 161219B, the value of $R_{\text{0,max}}$ is located around the center of the $R_0$ range, and as a result the peak energy is approximately constant.

For the G-Sample in Table \ref{tab:GStable} we have included the uncertainties that have been estimated in the detailed afterglow modeling of \cite{Laskar2013,Laskar2018c,Laskar2018,Laskar2019}. For the quantities with no errors provided, in order to estimate the uncertainties of our calculated photospheric quantities ($E_{\text{p}}$ and $E_{\text{p}}/E_{\text{p,obs}}$), we proceed by assuming typical uncertainty values based on the fractional uncertainties of the GRB quantities available.
For example, the only burst with a reported uncertainty in $\Gamma_{\text{dec}}$ is GRB181201A (see Table \ref{tab:GStable}), with a fractional error of $\sim 10$ per cent, so for the rest of the bursts we assumed the same level of uncertainty in $\Gamma_{\text{dec}}$. Similarly for $\sigma_{\text{dec}}$, we have scarce information about the uncertainties of $R_B$ and $\epsilon_{\text{B,f}}$, so we proceed by using a fixed fractional error of $50$ per cent to reflect the larger uncertainty of this quantity compared to $\Gamma_{dec}$.
We applied the same logic for the rest of the observed quantities ($E_{\text{iso}}, T_{90}, E_{\text{p,obs}}$) when uncertainties for these were not provided by the authors performing the afterglow modeling.


\subsection{Silver Sample: Known value of $\Gamma_{\text{dec}}$}

For our "Silver Sample" (S-Sample) we use data collected by \cite{Ghirlanda2018}. This sample has only a subset of the parameters known in comparison to the G-Sample: $(z,E_{\gamma,\text{iso}},L_{\gamma,\text{iso}},t_{\text{p}},\Gamma_{\text{dec}})$. These quantities are explained below. We used the values of the Lorentz factor at the deceleration radius $\Gamma_{\text{dec}}$ determined by \cite{Ghirlanda2018}, which they calculated by identifying the peak time of the afterglow light curve as the time that the blast wave starts decelerating. To do that, they used GRBs for which they have the afterglow peak time $t_{\text{p}}$, redshift $z$, and the gamma-ray isotropic energy $E_{\gamma,\text{iso}}$. 
\cite{Ghirlanda2018} calculated  $\Gamma_{\text{dec}}$ using \citep{Nava2013}:
\begin{equation}
    \Gamma_{\text{dec}}=\left[\frac{(17-4s)(9-2s)3^{2-s}}{2^{10-2s}\pi (4-s)}\left(\frac{E_{\text{iso}}}{n_0 m_{\text{p}} c^{5-s}}\right) \right]^{\frac{1}{8-2s}}t_{\text{p,z}}^{-\frac{3-s}{8-2s}},
\label{eq:Gdec}
\end{equation}
where $t_{\text{p,z}}=\frac{t_{\text{p}}}{1+z}$ is the redshifted afterglow peak time, and $E_{\text{iso}}=\frac{E_{\gamma,\text{iso}}}{\eta}$ is the total kinetic energy of the burst, assuming a prompt emission efficiency of $\eta=0.2$.

The medium density parameter $s$ is not known for this sample, so we can calculate the Lorentz factor for the two different density profiles 
$s=0$ and $s=2$. 
While our calculations for the S-Sample have been done for both values of $s$, for the rest of this paper we focus on the constant density medium case $s=0$. Afterglow modeling suggests that the majority of bursts are best fitted with a constant density medium (e.g., \citealp{Schulze2011}); a brief comparison of the expected photsopheric efficiencies and peak energies for a wind medium is presented in the discussion section. 

We will consider a neutron star (NS) and a black hole (BH) case for our central engine, with different values of $R_0$. We will follow two methods below. In the first method, we pick a characteristic (fixed) value of $R_0$. We will use the logarithmic mean value of $R_0$ for the corresponding range: $\bar{R}_{0,\text{NS}}=\frac{\text{max}[R_{0,NS}]-\text{min}[R_{0,NS}]}{\ln (\text{max}[R_{0,NS}]) - \ln (\text{min}[R_{0,NS}])}$ and $\bar{R}_{0,\text{BH}}=\frac{\text{max}[R_{0,BH}]-\text{min}[R_{0,BH}]}{\ln (\text{max}[R_{0,BH}]) - \ln (\text{min}[R_{0,BH}])}$. 
In the second method we will let $R_0$ vary between the minimum and maximum value for each case of the central engine. 

As mentioned before, we do not know the value of $\sigma_{\text{dec}}$ for the S-Sample, which is essential in the calculation of the stripe parameter $\alpha$. Instead, using the information from the G-Sample, we can assign a characteristic value for $\alpha$ for each value of $R_0$. As we can see in Fig. \ref{fig:alphaR0GS}, we have defined $\bar{\alpha}(R_0)$ as the average value of $\alpha$ for each value of $R_0$, based on the values of the parameter for each individual burst of the G-Sample. In the next sections we will see that small variations of $\alpha$ (like the range of $\bar{\alpha}$) do not have an important role in our calculations of the photospheric efficiency and the peak energies. As a result, we can define for the S-Sample $\bar{\alpha}_{S}\equiv\bar{\alpha}(\bar{R}_{0,BH})\approx 5/3$. This value will represent our S-sample for the two different cases of central engine, and that will allow us to proceed with the same calculations we did for the G-Sample. We could define a similar parameter for the NS case, but since $\alpha$ does not depend sensitively on $R_0$ we can use the same value for both NS and BH cases.

For the S-Sample, the uncertainty of the ratio $E_{\text{p}}/E_{\text{p,obs}}$ will depend on $\delta E_{\text{p}}$ and $\delta E_{\text{p,obs}}$. The former is obtained through error propagation of equation (\ref{eq:Epeak}) and will depend on the errors $\delta L_{\text{iso}},\delta \Gamma_{\text{dec}}$ and $\delta R_0$, while the latter observational error has been presented in \cite{Ghirlanda2018}. 
As done for the G-sample, we assume an error of $\sim 10$ per cent for $\Gamma_\text{dec}$ and we use $\delta L_{\text{iso}}$ and $\delta E_{\text{iso}}$ from \cite{Ghirlanda2018}. We assume $\delta R_{\text{0,NS}}=[\text{max}(R_{\text{0,NS}})-\text{min}(R_{\text{0,NS}})]/2$ and $\delta R_{\text{0,BH}}=[\text{max}(R_{\text{0,BH}})-\text{min}(R_{\text{0,BH}})]/2$. 

\subsubsection{Photospheric Efficiency of the Silver Sample bursts}\label{section:eff_phot}

\begin{figure*}
\centering
\begin{minipage}{.5\textwidth}
  \centering
  \includegraphics[width=1\linewidth]{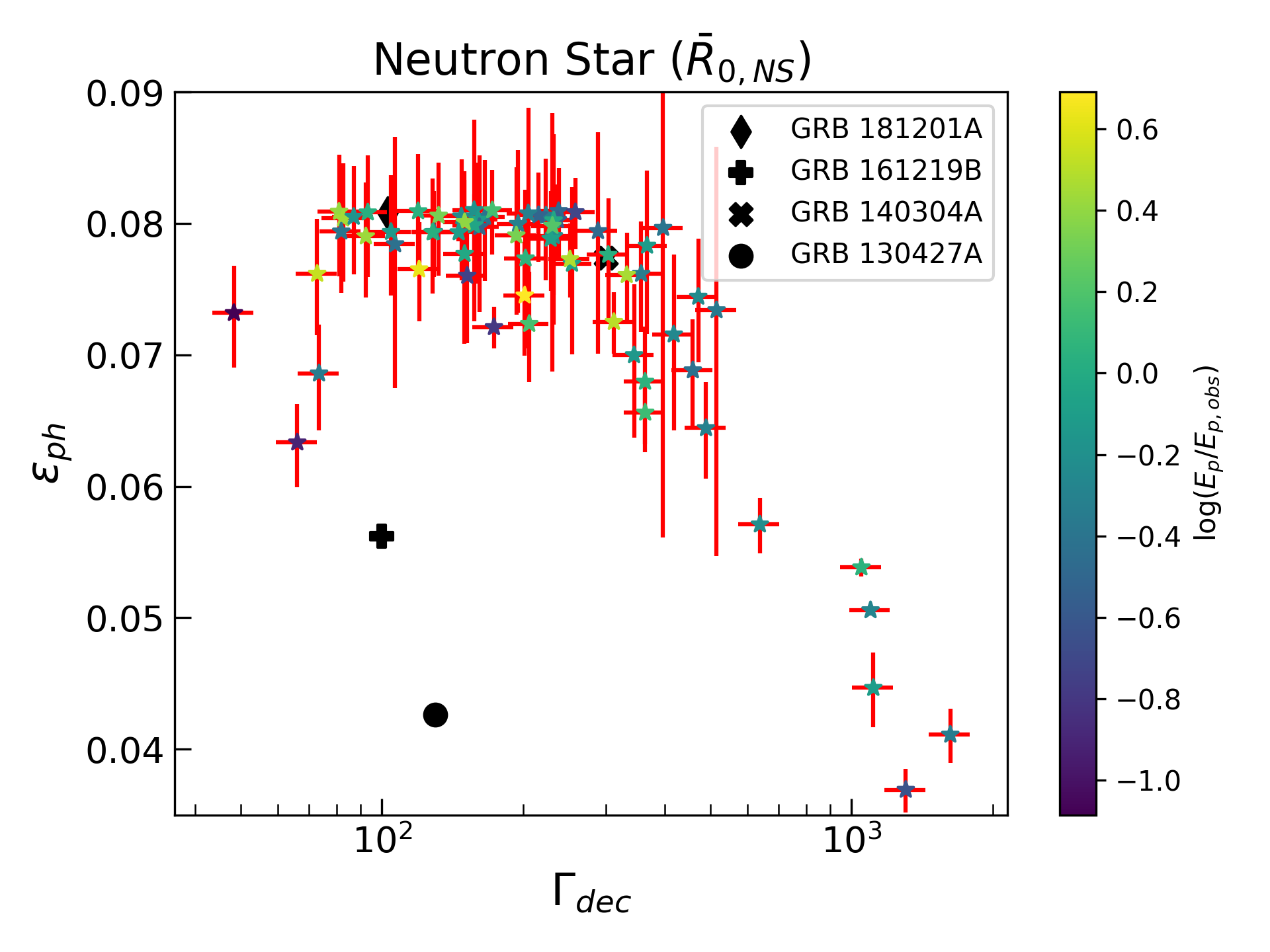}
\end{minipage}%
\begin{minipage}{.5\textwidth}
  \centering
  \includegraphics[width=1\linewidth]{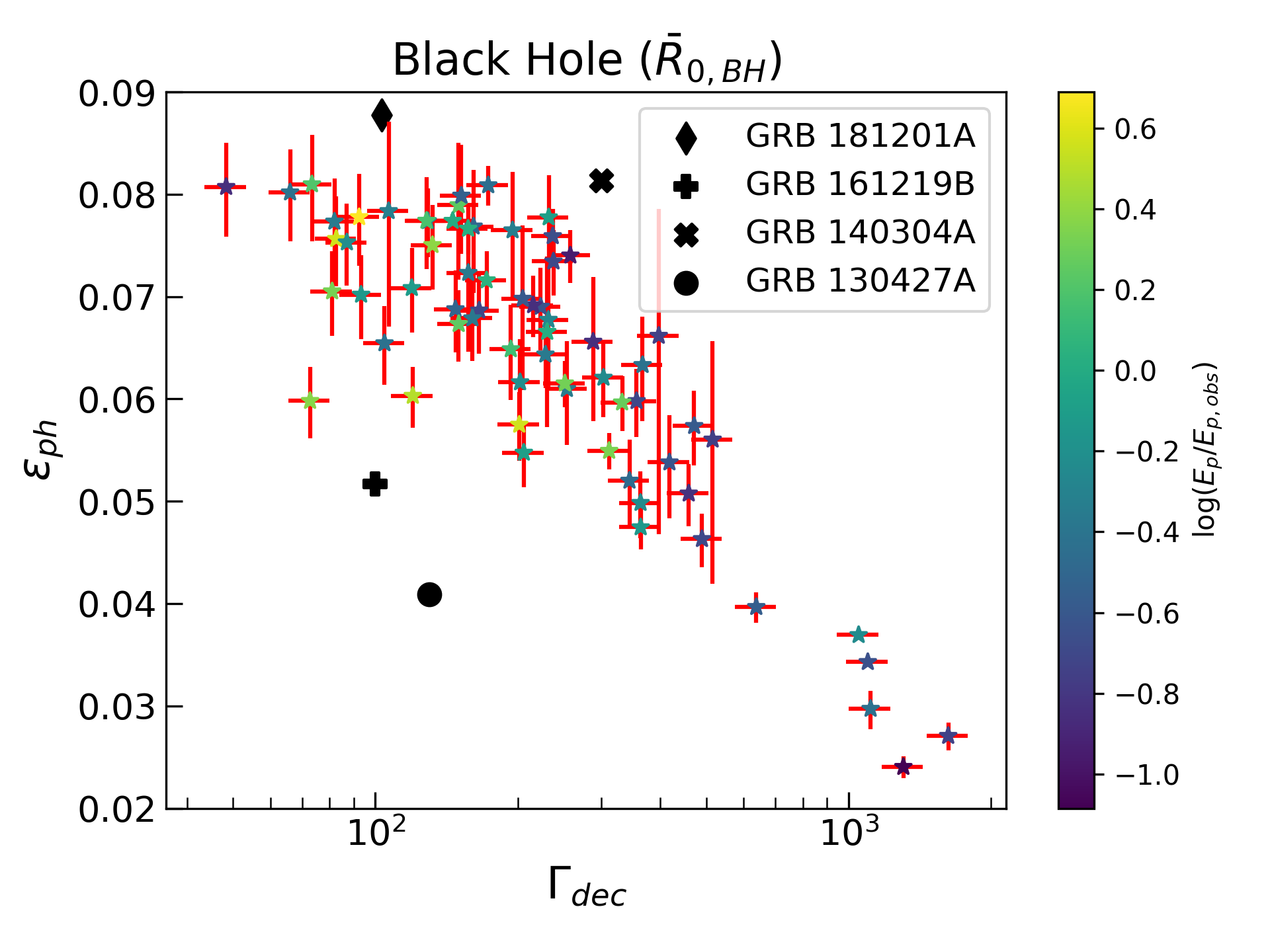}
\end{minipage}
\caption{Photospheric efficiency of the Silver Sample bursts for a neutron star (left, $R_0=\bar{R}_{\text{0,NS}}$) and a black hole (right, $R_0=\bar{R}_{\text{0,BH}}$) central engine. The Golden Sample bursts are also plotted separately in black (see legend). The color bar represents the logarithmic ratio $E_{\text{p}}/E_{\text{p,obs}}$. The maximum efficiency is almost the same in both cases, while the peak energies are more affected by the change of $R_0$.}
\label{fig:thermaleffSS}
\end{figure*}

\begin{figure}
	\includegraphics[width=\columnwidth]{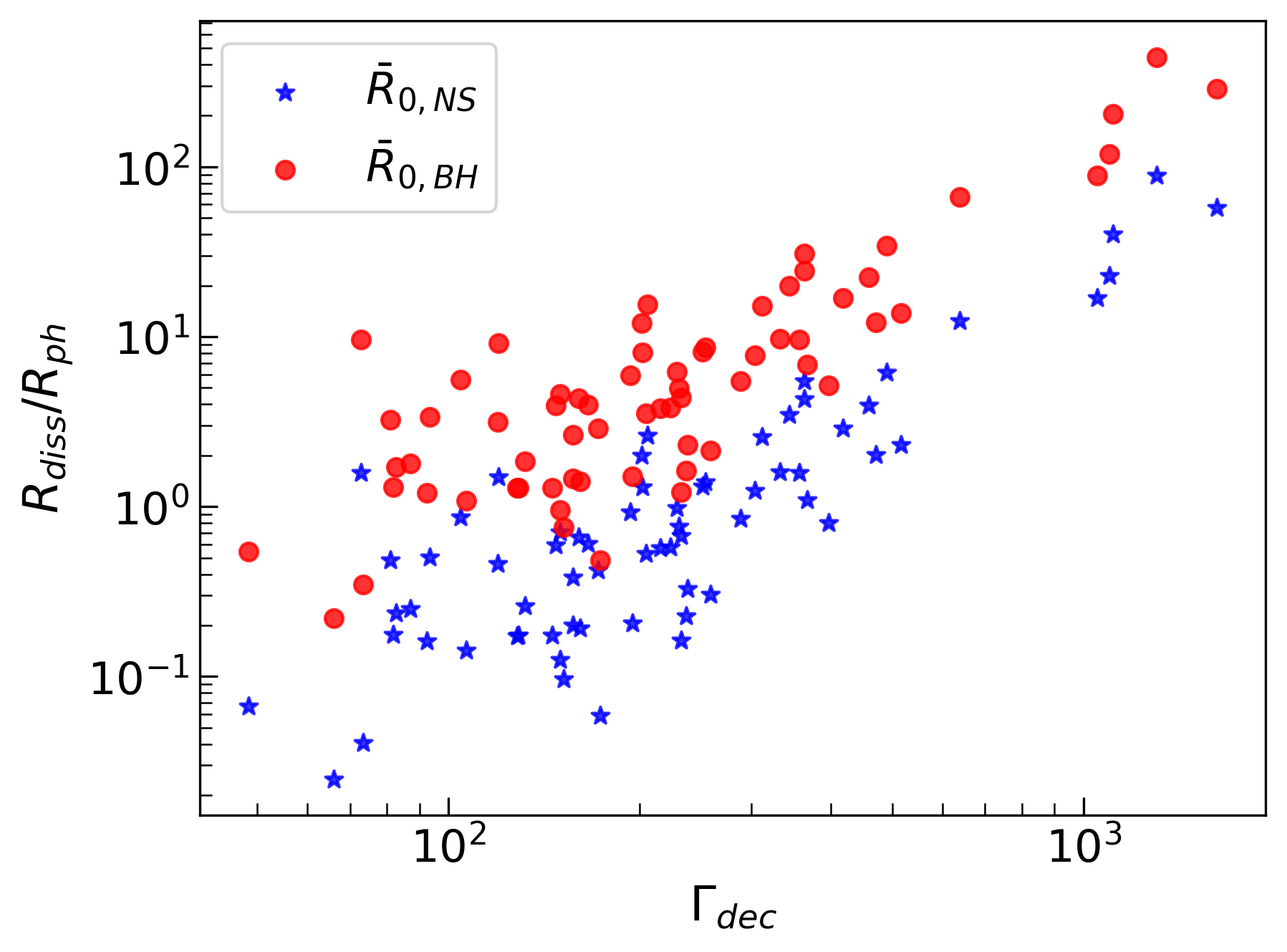}
    \caption{Ratio of the two characteristic radii $R_{\text{diss}}/R_{\text{ph}}$ for $R_0=\bar{R}_{\text{0,NS}}$ (blue) and $R_0=\bar{R}_{\text{0,BH}}$ (red). Changing the parameter $R_0$ from one for a NS to one of a BH the ratio of the two radii changes by about an order of magnitude, a change that does not affect much the photospheric efficiency of the bursts considering its $R_{\text{diss}}/R_{\text{ph}}$ efficiency in Fig. \ref{fig:thermaleff}. For a large number of bursts, the model predicts that the dissipation peaks close to the photospheric radius resulting in powerful photopsheric emission.}
    \label{fig:radiiratio}
\end{figure}

In Fig. \ref{fig:thermaleffSS} we plot the photospheric efficiency of both the S-Sample and the G-Sample\footnote{We note here that two of the G-Sample bursts (GRBs 181201A and 130427A) correspond to a wind external medium, in contrast to the other bursts of the sample and the S-Sample bursts where we have assumed a constant density medium. These two bursts serve only as a guideline since their efficiencies have been calculated for a wind external medium, while the rest of the points correspond to $s=0$.} for the two characteristic values of $R_0$: for neutron star ($\bar{R}_{\text{0,NS}}$ on the left) and for black hole ($\bar{R}_{0,BH}$ on the right). We see that the maximum value of the thermal efficiency, $\epsilon_{\text{ph,max}}$, in both cases is around $8$ per cent, almost unaffected by the different value of $R_0$, contrary to the peak energy which depends more sensitively on $R_0$ (we used the same color bar scale for both plots to see the difference). 
Also, we notice the existence of an efficiency "plateau", close to $\epsilon_{\text{ph,max}}$ for the range of small $\Gamma_{\text{dec}}$ appreciated more in the left plot of Fig. \ref{fig:thermaleffSS}. This is because we can see in Fig. \ref{fig:thermaleff} that a) the maximum efficiency occurs when $R_{\text{diss}}$ and $R_{\text{ph}}$ are comparable, and b) that the smaller the value of $\alpha$ is, the less sensitive is the dependence of the efficiency on $R_{\text{diss}}/R_{\text{ph}}$ (and consequently the dependence on $R_0$) around the maximum efficiency. 
Indeed, as we can see from Fig. \ref{fig:radiiratio}, changing the central engine from a neutron star to a black hole results in a change of $R_{\text{diss}}/R_{\text{ph}}$ by less than an order of magnitude. Also, for the range of small $\Gamma_{\text{dec}}$ mentioned above, that ratio in both cases is $\sim 0.1-10$. As a result, for a large subset of our sample the efficiency will be close to the maximum value. 
Even though we initially assumed a value $\eta=0.2$ for the gamma-ray efficiency of the bursts, our calculated value of $\sim 0.08$ is still consistent with our assumptions and corresponds to bursts with a significant photospheric component in gamma-rays. Repeating our calculations with the new value of $\eta=0.08$, we find that even though the deceleration radius quantities $R_{\text{dec}}$ and $\Gamma_{\text{dec}}$ change by around $20$ and $10$ per cent, respectively, due to the dependence of $\epsilon_{\text{ph}}$ on $R_{\text{diss}}/R_{\text{ph}}$ only, the photospheric efficiencies remain almost unaffected by the change ($\sim$ 2-3\% change of the original values). 

\subsubsection{The nature of the central engine: Constant value of $R_0$ for all bursts}
\label{sec:oneR0}

\begin{figure*}
\centering
\begin{minipage}{.5\textwidth}
  \centering
  \includegraphics[width=1\linewidth]{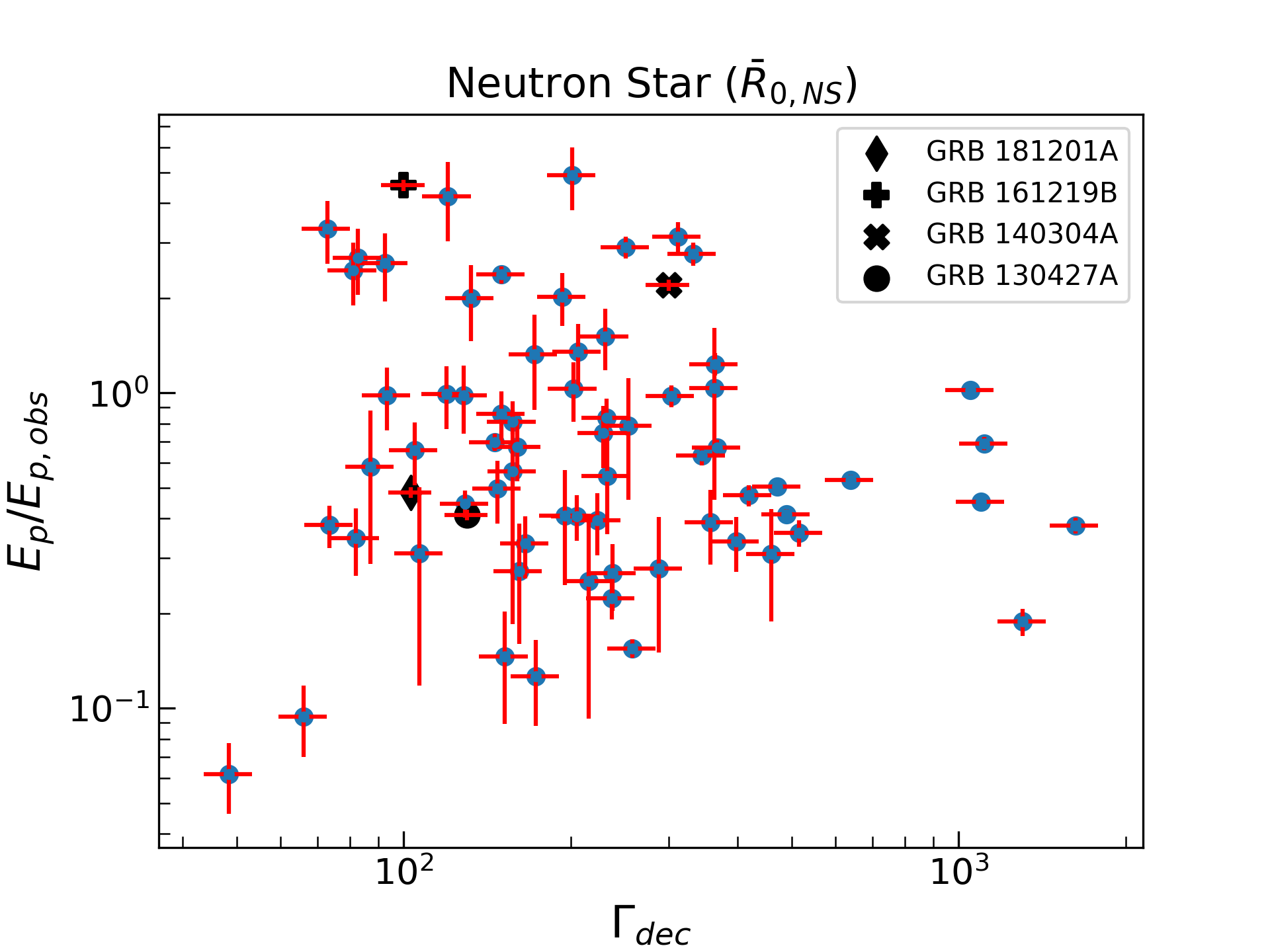}
\end{minipage}%
\begin{minipage}{.5\textwidth}
  \centering
  \includegraphics[width=1\linewidth]{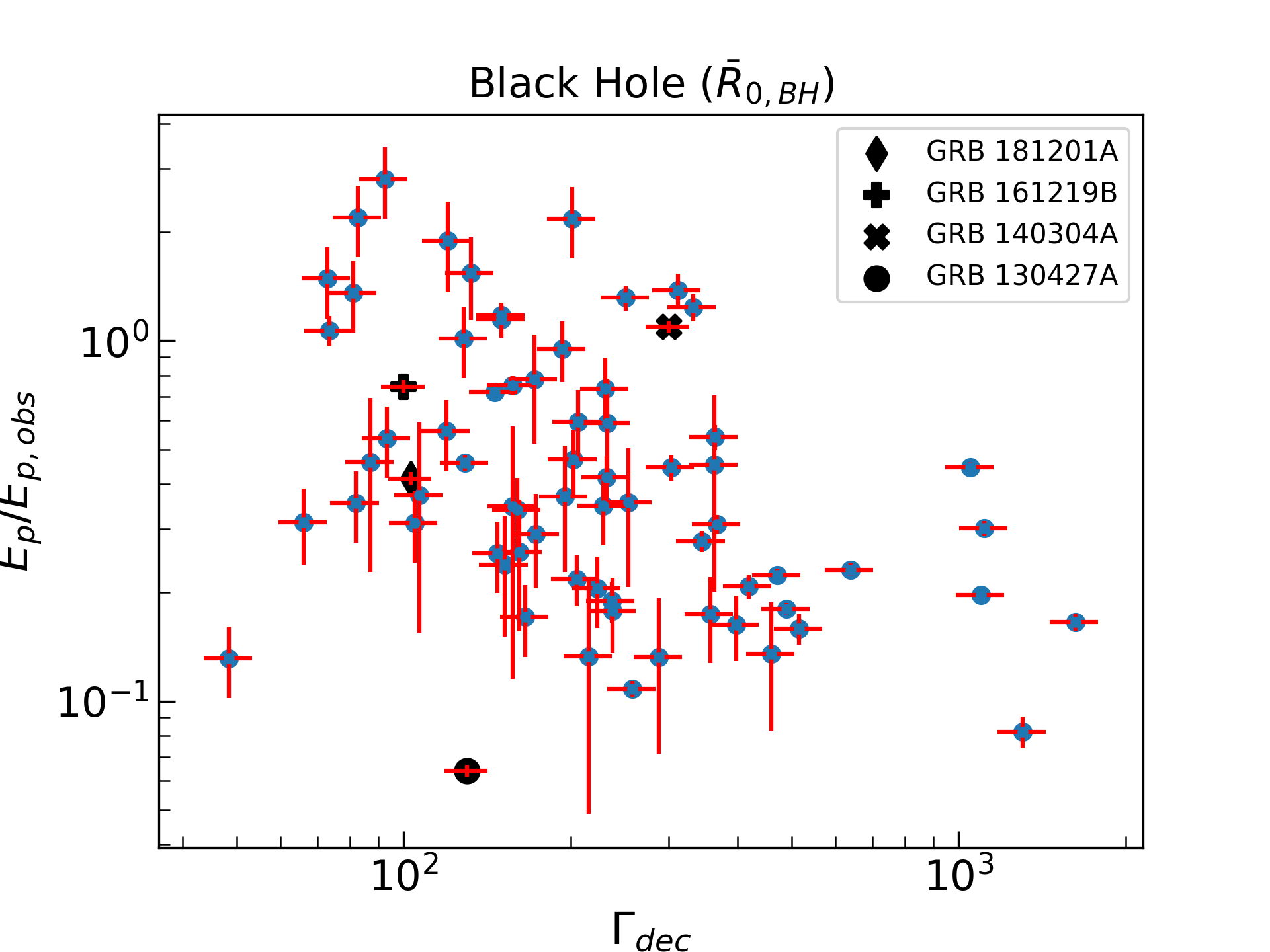}
\end{minipage}
\caption{Ratio of the theoretical to the observed peak energy as a function of the Lorentz factor at the deceleration radius for a neutron star (left, $R_0=\bar{R}_{\text{0,NS}}$) and a black hole (right, $R_0=\bar{R}_{\text{0,BH}}$) central engine. We can see that on average, the peak energy drops for the BH case, where $R_0$ is larger. We also notice that the faster bursts ($\Gamma_{\text{dec}}\gtrsim 1000$) have a very small error on the y axis. That is because the faster bursts are also the more luminous ones, so they have a very clear observation of the peak energy.}
\label{fig:EpratioSS}
\end{figure*}

In Fig. \ref{fig:EpratioSS} we plot the ratio $E_{\text{p}}/E_{\text{p,obs}}$ for the two characteristic values of $R_0$ for NS and BH, respectively. We find that the S-Sample in the case of a NS engine has more bursts with $E_{\text{p}}/E_{\text{p,obs}} \sim 1$ compared to the BH case. More specifically, the NS case has 28 (42 per cent) bursts with  $E_{\text{p}}/E_{\text{p,obs}} = 1$ within a factor of 2, and 58 (87 per cent) bursts within a factor of 4. On the other hand, the BH case has 21 (31 per cent) and 46 (69 per cent) bursts with $E_{\text{p}}/E_{\text{p,obs}} = 1$ within a factor of 2 and 4, respectively. 

\begin{figure}
	\includegraphics[width=\columnwidth]{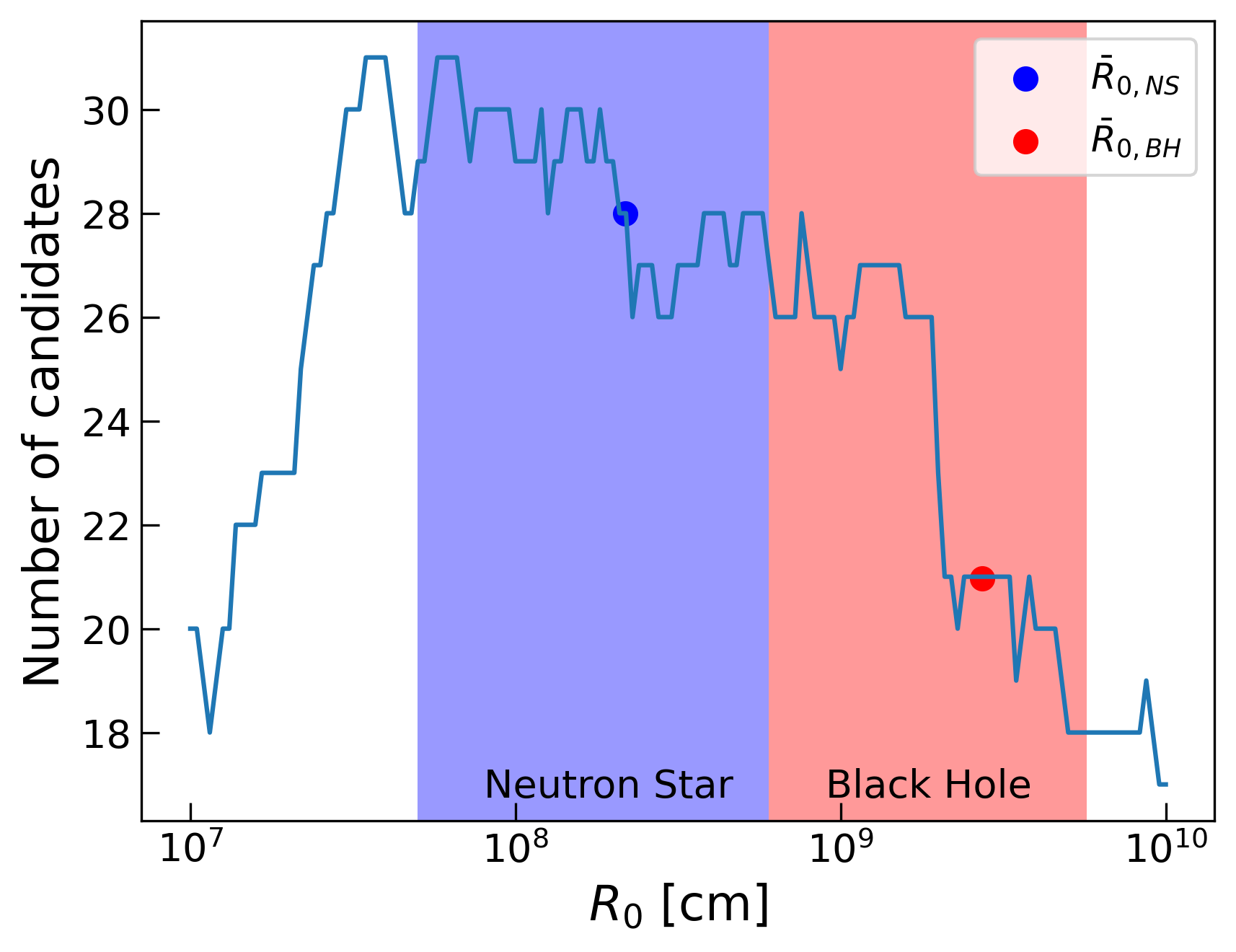}
    \caption{Number of S-Sample bursts that have $E_{\text{p}}$ within a factor of 2 from $E_{\text{p,obs}}$ for $R_0$ in the range $10^7-10^{10}$ cm. The blue and red dots correspond to the ratios that have been calculated in Fig. \ref{fig:EpratioSS} for $R_0=\bar{R}_{\text{0,NS}}$ and $R_0=\bar{R}_{\text{0,BH}}$, respectively. More bursts have a peak energy compatible with a neutron star central engine.}
    \label{fig:R0frac}
\end{figure}

We can expand our results and estimate how well our calculated peak energies match the observed ones for all values of $R_0$ within the range of the NS and BH central engines. 
We essentially repeat Fig. \ref{fig:EpratioSS} for the entire range of $R_0$ values, from $\text{min}(R_{\text{0,NS}})$ to $\text{max}(R_{\text{0,BH}})$, and for each value of $R_0$ we calculate what fraction of the S-Sample bursts have a theoretical peak energy $E_\text{p}(R_0)$ within a factor of 2 from the observed value $E_{\text{p,obs}}$. We present the results in Fig. \ref{fig:R0frac}. We see that this method favors neutron stars as central engines, with an average fraction of accepted bursts of $\sim 43$ per cent (29 out of 67 bursts), compared to the black hole case with an average fraction of $\sim 34$ per cent (23 out of 67 bursts). 

We should note here that the 
number of candidates on this plot are {\it potential} candidates for a NS/BH central engine. In other words, based on the $R_0$ dependence of the peak energy, there could be bursts that satisfy the conditions for both a NS and a BH central engine, thus they are going to contribute to the fractions in both blue and red areas in Fig. \ref{fig:R0frac}. This leads to the method presented in the next subsection. 

\subsubsection{The nature of the central engine: Varying the value of $R_0$}

Though we seem to find more neutron star candidates in Section \ref{sec:oneR0}, the downside of this method is that we assign a specific value of $R_0$ to all of the bursts. This is a a convenient way to simplify our results by reducing the number of parameters to one, but one value of $R_0$ cannot accurately represent all of our sample. Ideally, we would like to find a different value of $R_0$ for each burst based on how well it fits the observational data, and probe the nature of the compact object favored by the model as a central engine candidate. 

\begin{figure}
	\includegraphics[width=\columnwidth]{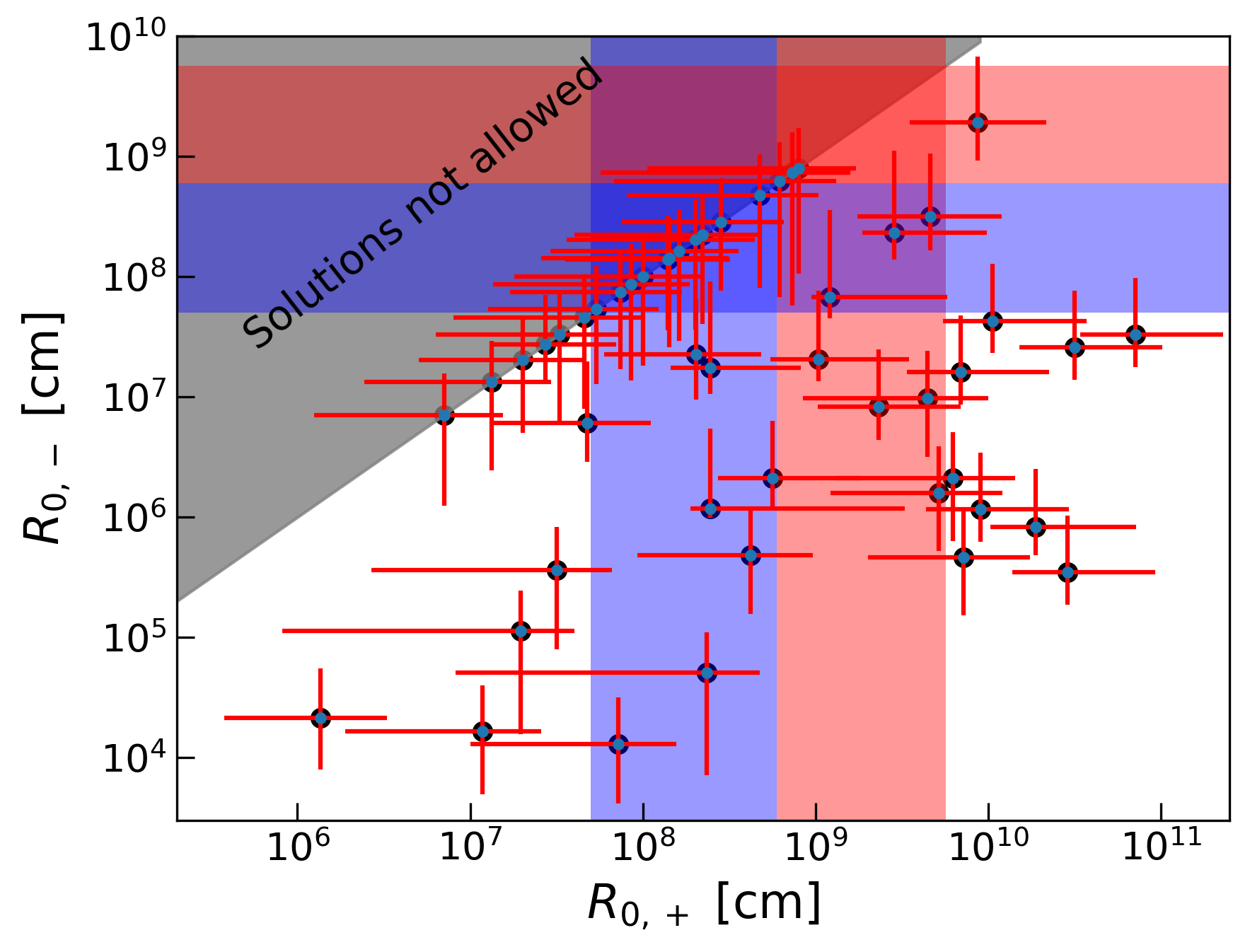}
    \caption{Scatter plot of the acceptable solutions for $R_0$. The x-axis ($R_{0,+}$) corresponds to the largest of the two solutions, while the y-axis ($R_{0,-}$) corresponds to the smaller one. The blue and red areas correspond to the acceptable regions for NS and BH central engine,respectively. By definition, no solutions are allowed in the gray area above the $y=x$ line.}
\label{fig:R0solutions}
\end{figure}

For this reason, we now consider another approach to find what central engine is favored by our model, which is orthogonal to the first one. Instead of keeping $R_0$ constant, we now vary $R_0$ for each burst, in a way that yields a peak energy $E_{\text{p}}$ that is as close as possible to the observed value $E_{\text{p,obs}}$. We can then analyze the distribution of the values of $R_0$ (for example through a histogram) to find the one with the highest frequency. Specifically, for each burst of the S-Sample, 
we solve the equation 
$E_{\text{p}}(R_0)=E_{\text{p,obs}}$ for $R_0$ based on equation (\ref{eq:Epeak}). 
As we mentioned before, the dependence of $E_{\text{p}}$ on $R_0$ is not monotonic, and up to two values of $R_0$ can give the same value for the peak energy, depending on the value of $R_0$ compared to $R_{\text{0,max}}$. 
As a result, we can divide the burst classification into two different cases: If for a burst 
we find $E_{\text{p,obs}}\le E_{\text{p,max}}$, then there will be two solutions for $R_0$ that satisfy $E_{\text{p}}=E_{\text{p,obs}}$, which we call $R_{0,\pm}$ based on which one is larger/smaller than $R_{\text{0,max}}$. On the other hand, if $E_{\text{p,obs}}>E_{\text{p,max}}$, then the equation above has no solutions for $R_0$, which means that the model cannot accurately predict the observed value of the peak energy. In those cases (including also the cases where $E_{\text{p,obs}}=E_{\text{p,max}}$), we choose to keep the value $R_{\text{0,max}}$ as the characteristic value of $R_0$, since it is the value that gives the closest value of $E_{\text{p}}$ compared to $E_{\text{p,obs}}$. We present the results of this calculation as histograms of $R_{0,+}$ and $R_{0,-}$ in Fig. \ref{fig:R0solutions}. 

Similarly to the first method, we are only examining the bursts that give a peak energy within a factor of 2 from the observed value (50 out of 67 bursts, 75 per cent of the S-Sample), and the errors on $R_{0,\pm}$ have been calculated based on the uncertainties 
in $L_{\text{iso}}$, $\Gamma_{\text{dec}}$, $E_{\text{p,obs}}$ as described before in this Section. We notice, based on the calculated values of $R_{0,\pm}$ and the ranges of $R_0$ we have considered for NS and BH, that:

\begin{itemize}
    \item For 19 bursts, either both $R_{0,\pm}$ are in the neutron star area, or $R_{0,+}$ is compatible with a NS and $R_{0,-}$ has an unrealistic value. As a result, 19 bursts in total are compatible only with a neutron star central engine. 
    \item Similarly, 8 bursts are compatible only with a black hole central engine. 
    \item 15 bursts are candidates for both a neutron star and a black hole central engine.
    \item 8 bursts have incompatible values of $R_{0,\pm}$ for both NS and BH central engines; their values are either too small or too large for our assigned range of $R_0$.
    \item Finally as mentioned before, 17 bursts were excluded from the sample because even though they could give a realistic value of $R_0$, their theoretical peak energy differs significantly from the observed value, and as a result they are not compatible with our model. 
\end{itemize}

\begin{figure}
	\includegraphics[width=\columnwidth]{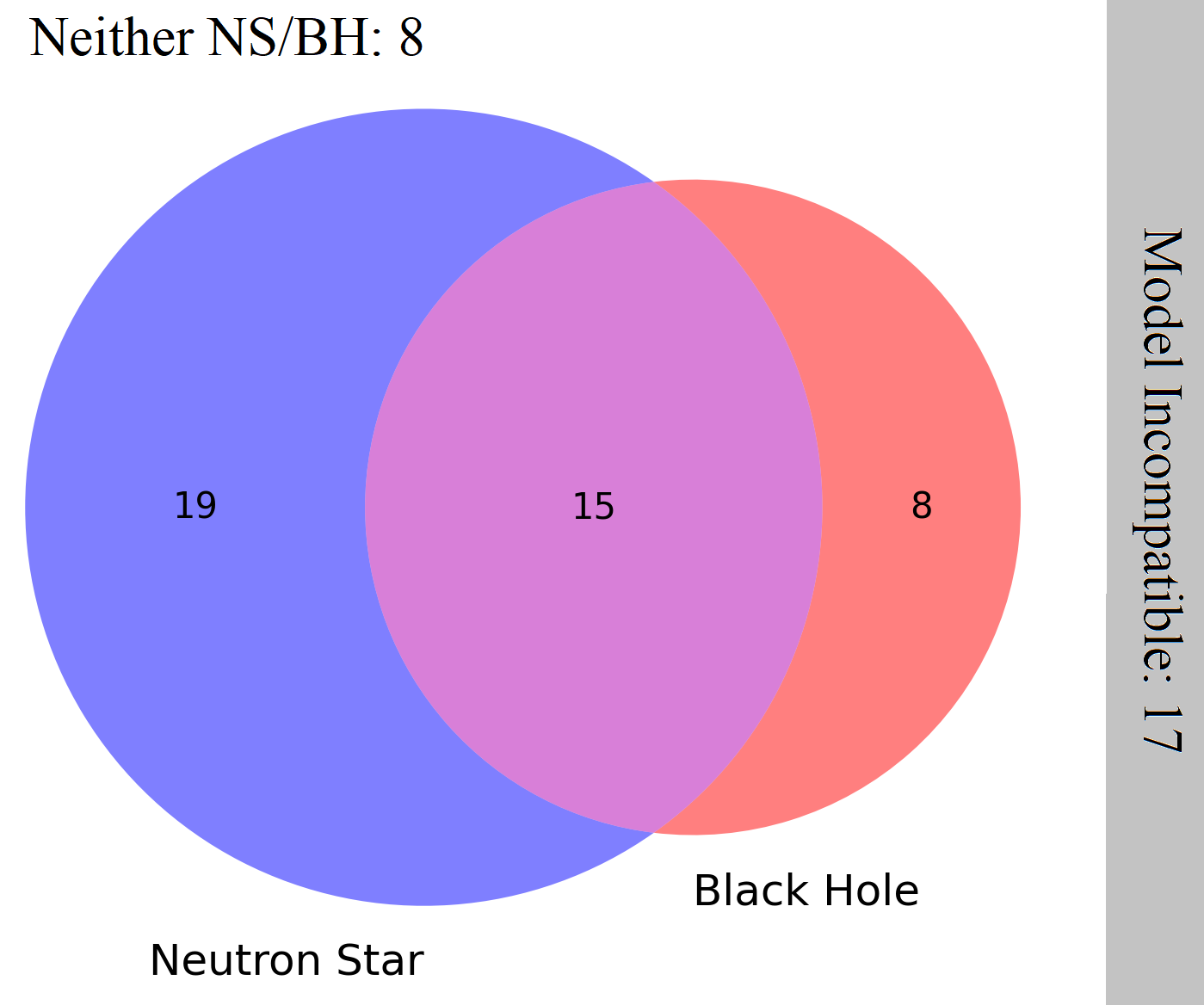}
    \caption{Venn diagram of the neutron star and black hole inferred engine sets. There are 19 neutron star candidates, 8 black hole candidates, and 15 bursts that are potential candidates for both NS and BH central engines. Furthermore, 8 bursts have unrealistic values of $R_0$ for both kinds of central engines, and 17 bursts were excluded from the sample due to unrealistic values of the peak energy.}
    \label{fig:venn}
\end{figure}

These results have been organized in a Venn diagram in Figure \ref{fig:venn}.

\section{Discussion}
In this work we are using GRB observations of the prompt emission ($L_{\text{iso}}$) and the afterglow ($\sigma_{\text{dec}}$, $\Gamma_{\infty}$) to connect these two distinct processes of the burst at the deceleration radius ($\sim10^{17}$cm), and infer more information about the nature of the central engine favored by our striped jet model. The importance of this analysis is that by knowing the burst's parameters at a large distance ($r=R_{\text{dec}}$) we can (almost) uniquely determine the jet's properties at smaller distances, such as the acceleration profile $\Gamma(r)$, the magnetization $\sigma(r)$ and the stripe distribution $\alpha$. 

We find the photospheric efficiencies and the peak energies of the bursts (with the former being insensitive to the change of central engine and density profile of the external medium, and the latter being affected more by the change of parameters $R_0$ and $s$) and our study points towards a neutron star engine 
being slightly more favorable by our model. The number of free parameters used in this study is very small: we are changing $R_0$ and $\alpha$, with the latter being fixed for the most part of this work. We use a constant density medium 
($s=0$), which is consistent with previous works analyzing large number of afterglows 
(e.g., \citealp{Schulze2011}). 
By analyzing the largest of our samples (S-Sample, 67 bursts) we calculate the photospheric efficiencies of the bursts. We find their values in a narrow range of $\sim5-10$ per cent, with a maximum efficiency almost independent of the choice of $R_0$ due to the dependence of $\epsilon_{\text{ph}}$ on $R_{\text{diss}}/R_{\text{ph}}$. 
Similarly, the study of the thermal emission below the photosphere for the G-Sample (Section \ref{section:GoldenSample}) showed a weak dependence of the photospheric efficiency on $R_0$, with values again between $\sim 5-10$ per cent depending on the burst.

The photospheric component of the emission does not correspond to a blackbody, rather a modified spectrum with a flat high-energy tail due to the Comptonization of the thermal photons close to the photospheric radius (\citealp{Giannios2006}).  
We also notice that the bursts with the highest asymptotic Lorentz factors ($\Gamma_{\text{dec}}\gtrsim 1000$) have lower photospheric efficiencies compared to the rest of the sample. At the same time, these bursts also have the highest upper limit for the non-thermal luminosity and efficiency ($\sim 40\%$). We might expect for these bursts to have a stronger contribution to the total luminosity from non-thermal components since the photospheric component is weak. 
This is consistent with Fermi observations of fast Gamma-Ray bursts like GRB090902B and GRB090926A, where time-resolved spectral analysis shows the existence of an additional power-law component in the spectrum \citep{090902B,090926A}. 
We note here that if the photospheric efficiency is assumed to be equal to the efficiency of the prompt emission, this could lead in some limited cases to an underestimation of the total efficiency. As explained below, the fastest of the bursts (which correspond to $R_{\text{diss}}\gg R_{\text{ph}}$) also have the lowest photospheric efficiencies in our S-Sample. That could be explained by strong optically thin emission which we do not attempt to model here. Our assumption that the prompt emission efficiency is dominated by the photospheric component is valid for the rest of the bursts, something that is also supported by previous work (\citealp{Giannios2006}).

\begin{figure}
\label{fig:}
	\includegraphics[width=\columnwidth]{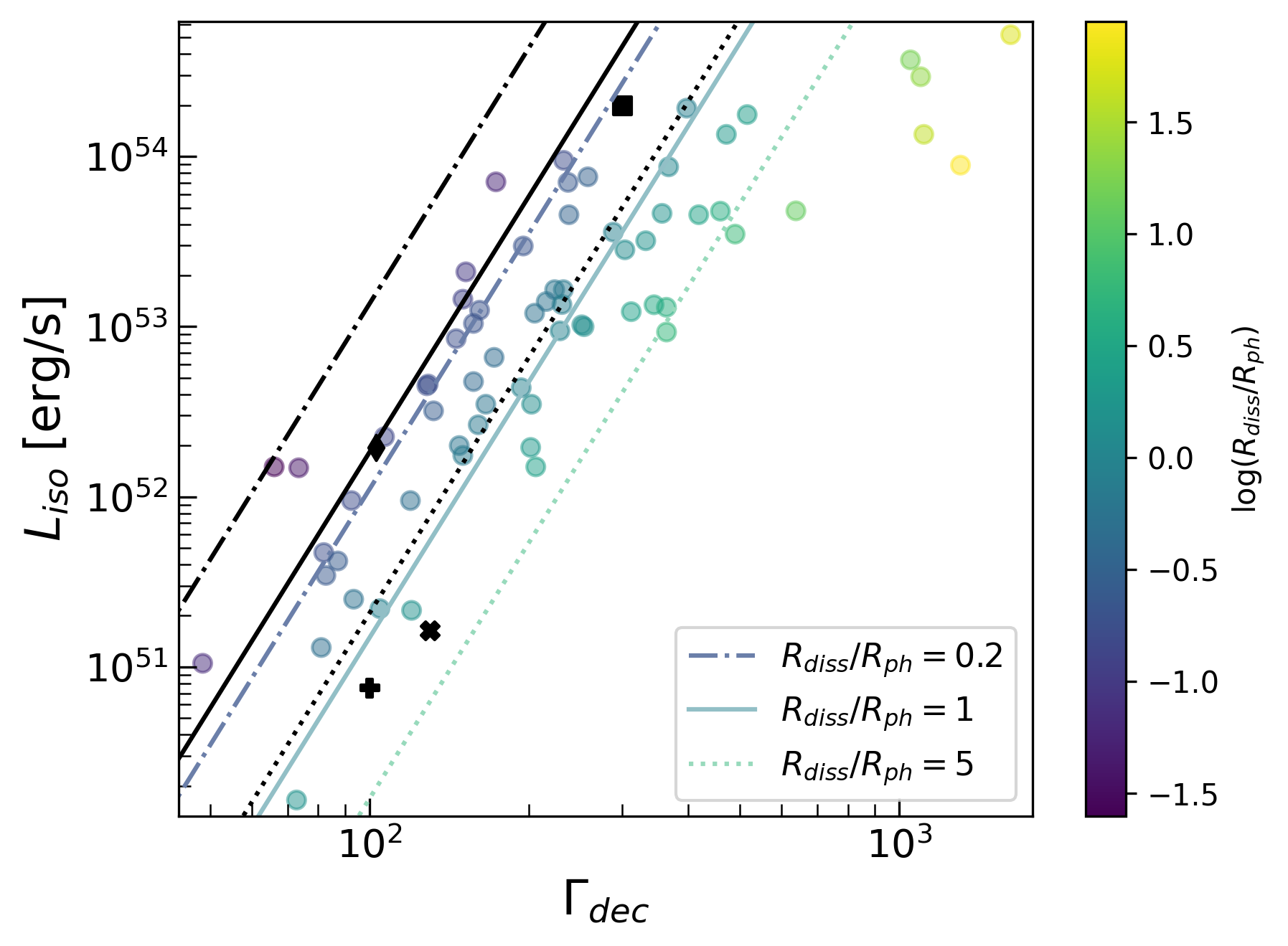}
    \caption{Plot of $L_{\text{iso}}$ and $\Gamma_{\text{dec}}$ of the Golden and Silver samples for a constant density medium.
    The straight lines show a $L_{\text{iso}}\propto \Gamma_{\text{dec}}^5$ dependence for different values of $R_{\text{diss}}/R_{\text{ph}}$, as inferred from equation (\ref{eq:LGdependence}), while the color bar shows the full range of the ratio of the two radii for $R_0=\bar{R}_{\text{0,NS}}$.
    The colored lines correspond to $\bar{R}_{\text{0,NS}}$ while the corresponding black lines to $\bar{R}_{\text{0,BH}}$ for the same values of the ratio of the two radii. For a large fraction of the bursts, the model predicts that $R_{\text{diss}}\sim R_{\text{ph}}$.}
    \label{fig:LGdec}
\end{figure}

In Fig. \ref{fig:LGdec} we plot the correlation between isotropic luminosities and Lorentz factors at the deceleration radius for each burst of the Golden and Silver samples, accompanied by their corresponding value of $\frac{R_{\text{diss}}}{R_{\text{ph}}}$ for $R_0=\bar{R}_{\text{0,NS}}$. The continuous and dashed lines correspond to a $L_{\text{iso}}\propto \Gamma_{\text{dec}}^5$ power-law dependence for $R_{\text{diss}}/R_{\text{ph}}=\text{const}$, since from equation (\ref{eq:opticaldepth}) at the photospheric radius we have:
\begin{equation}
    L_{\text{iso}}=\frac{24\pi c^3 R_0}{\kappa}\left(\int_{\chi_{\text{ph}}}^{1}\frac{d\chi'}{(1-\chi')^k\zeta^2_k(\chi')}\right)^{-1}\Gamma_{\infty}^5.
    \label{eq:LGdependence}
\end{equation}
The proportionality constant depends strongly on $R_0$ and the ratio of the two radii, and also weakly on $\alpha$. It is very interesting that
by varying $R_{\text{diss}}/R_{\text{ph}}$ by a factor of 5 from unity, and $R_0$ in the NS/BH range, most bursts fall into this range of power-laws. The exception is bursts with very high Lorentz factors ($\Gamma_{\text{dec}}\gtrsim 1000$) as mentioned above. These bursts correspond to a much larger ratio $R_{\text{diss}}/R_{\text{ph}}\sim 10^2$ and as a result to a relatively small photospheric radiative efficiency. The model predicts that these bursts have a stronger contribution to the total luminosity from a non-thermal component (e.g., \citealp{Gill2020}). In fact, it is likely that these bursts dissipate their energy via `minijets', i.e., regions which move relativistically relative to the main flow of the jet result of magnetic reconnection of the stripes (e.g., \citealp{Barniolduran2016, Beniamini2016}).  It is interesting that a correlation between prompt variability and prompt peak luminosity exists (e.g., \citealp{Guidorzi2005}, see \citealp{Beniamini2016} for a discussion of this correlation in the context of minijets). This may point to the fact that these more luminous bursts can be more variable, consistent with a minijet origin with a strong non-thermal component as found in this study.

For our analysis we have assumed a stripe parameter consistent with the G-Sample data ($\alpha=\bar{\alpha}(\bar{R}_{0,\text{BH}})\approx5/3$) so we can proceed with our calculations. This is due to the lack of sufficient afterglow observations for the majority of the GRBs, that results in poor/nonexistent constraints of the jet magnetization at the deceleration radius. It is an estimate of the jet magnetization in the G-Sample which allows for a determination of $\alpha$ for a small number of bursts. Considering the fact that our G-Sample is small, the average value of $\alpha$ could possibly vary for a larger sample of bursts (such as the S-Sample) with good constrains of the jet magnetization. The varying value of $\alpha$ could be a result of a broad range of values for $E_{\text{iso}}$ and $\Gamma_{\text{dec}}$ (which is the case for the S-Sample), as well as the uncertainty of $R_0$ and the potentially broad range of the magnetization $\sigma_{\text{dec}}$. Repeating the analysis of section \ref{section:eff_phot}, and calculating the photospheric efficiency of the S-Sample for a large value of $\alpha=10$, we find the values of $\epsilon_{\text{ph}}$ to be slightly larger, with a maximum efficiency of $\sim 10\%$ instead of $\sim 8\%$, in agreement with Fig. \ref{fig:thermaleff}. For both cases of $\alpha$ the values are within the expected for photospheric emission, and since the dependence on that parameter is relatively weak, we only use one value of $\alpha$ for our calculations.

We use the striped jet model to calculate the theoretical photospheric peak energies for our Gold and Silver Samples. We find that for two of the G-Sample bursts (GRB181201A and GRB161219B) the $\Gamma_{\text{dec}}$ and $L_{\text{iso}}$ parameters are such that the range of $R_0$ we have considered is around the maximum of the function $E_{\text{p}}(R_0)$, while for the other two (GRB140304A and GRB130427A) we have $R_0>R_{0,\text{max}}$ for all values of $R_0$ in our range. For this reason, for the former two we cannot deduce the optimal range of $R_0$ or the type of central engine, as most values of $R_0$ give approximately the same value of $E_{\text{p}}/E_{\text{p,obs}}$, which is remarkably close to $E_{\text{p}}\simeq E_{\text{p,obs}}$. On the other hand, the second two bursts have a monotonic, decreasing dependence on $R_0$, and as we can see from Fig. \ref{fig:EpeakratioGS}, the photospheric model agrees more with a BH central engine for GRB140304A, and a NS engine for GRB130427A.

For our S-Sample we calculate the photospheric peak energies with $R_0$ as a free parameter. Due to the large sample size, we use two different methods to determine what values of $R_0$ are favored by our model and the observational data: one that keeps $R_0$ constant for all the bursts and calculates $E_{\text{p}}$, and one that estimates the value of $R_0$ by minimizing the quantity ($E_{\text{p}}/E_{\text{p,obs}}-1$). 
The first method shows our model favoring a neutron star central engine ($\sim43$ per cent of the bursts are candidates) over a black hole one ($\sim34$ per cent), and the number of potential candidates for both cases to decrease as a function of $R_0$ (Fig. \ref{fig:R0frac}). The second method takes into account the $R_0$ degeneracy, where one value of $E_{\text{p}}$ corresponds to two values of $R_0$, one where the dissipation peaks above and one below the photospheric radius, respectively. This second method shows that the striped jet model yields peak energies consistent with the observed values and realistic $R_0$ values for about 2/3 of the bursts in our sample. By calculating both solutions that give a peak energy as close as possible to the observed value, we find 19 GRBs to be compatible only with a NS central engine compared to 8 bursts for a BH, thus favoring again a magnetar as the central engine for a larger number of GRBs. 

We note here that our calculation of the peak energy is based on the assumption that there is complete thermalization at high optical depths. This implies that the photon production mechanisms deep inside the Thomson photosphere are efficient enough to produce a Planck spectrum. Ultimately, the validity of this assumption will depend on the particle acceleration mechanisms operating in these regions, something that remains highly uncertain. If the thermalization process is not complete, we would instead obtain higher peak energies for our sample (\citealp{Vurm2013, Begue2015}), which could alleviate the problem of low peak energies in a modest fraction of our sample (17 out of 67 bursts).

As mentioned earlier, for the majority of this work we assumed a constant density medium $(s=0)$ to calculate the asymptotic Lorentz factors and the deceleration radius. 
Repeating the analysis for a wind case external medium $(s=2)$ we find the change to have a very small effect in the range and maximum value of the photospheric efficiency. The reason for this is that a change of the wind profile leads to a decrease of the asymptotic Lorentz factor, and consequently from equation (\ref{eq:RdissRph}), to a decrease in $R_{\text{diss}}/R_{\text{ph}}$. As a result, the efficiency-$\Gamma_{\text{dec}}$ dependence in Fig. \ref{fig:thermaleffSS} shifts to the left, and bursts of high/low Lorentz factor have slightly larger/lower efficiencies, without affecting $\epsilon_{\text{ph,max}}$.
On the other hand, the change of the density profile has a significant effect on the peak energies. For the wind case, the number of bursts that have theoretical peak energies within a factor of 2 from the observed values drops by a factor of 3 compared to the constant density case. The drop in peak energy comes from equations (\ref{eq:Epeak}) and (\ref{eq:Gdec}), where the Lorentz factor at the deceleration radius drops for a wind medium, and so does the peak energy. 

Another goal of our work was to estimate the stripe distribution of the jet's magnetic field from its magnetization at the deceleration radius. By analyzing the G-Sample data, and using the magnetization of the ejecta we calculate the dimensionless quantities $\chi_{\text{dec}}$ and $\zeta_{\text{dec}}(R_0)$. Using those and the methodology described on section \ref{section:GoldenSample}, we can obtain an estimate of the $\alpha$ parameter (and thus the stripe distribution of the jet) for each burst. Since the distance $\zeta$ is a function of $R_0$, our results for the stripe parameter and every quantity that depends on it will also depend on $R_0$. We find that the G-Sample data combined with our model favors small values of $\alpha$ ($\alpha<2$), and that larger values of $R_0$ also give larger values of $\alpha$. The values of $\alpha$ that we find correspond to models with a broader dissipation profile, where even though the dissipation peaks at $R_{\text{diss}}$, it proceeds on multiple length-scales. This is contrary to the model of \cite{Drenkhahn2002}, where the dissipation profile is steep and the value of $\alpha$ very large. A value of $\alpha \sim 5/3$, compatible with observations, has been argued by \cite{Giannios_Uzdensky2019}. In that work, it was argued that the generated stripe time (and length) scales are proportional to the Keplerian period in the accretion disk and that the generated power scales with the local gravitational energy that is released. That value of $\alpha$ could be more applicable for black holes or neutron stars surrounded by an accretion disk. For the former case, an analysis of GRBs with X-ray afterglow plateaus (\citealp{Dainotti2017,Stratta2018}) provides support to accreting magnetars as the central engine, which would correspond to a value of $\alpha$ close to our estimations. Ultimately, 3D GR-MHD simulations (\citealp{Christie2019}) can provide more insight about the magnetic field geometry in such situations.

Finally, we note that our results depend on the Lorentz factor estimates performed in \cite{Ghirlanda2018}. They determine the asymptotic Lorentz factor of the jet assuming that the afterglow peak time $t_{\text{p}}$ corresponds to the deceleration time of the blast wave, and also that the reverse shock is Newtonian. The latter has been used in equation (\ref{eq:sigma}) in order to find the magnetization of the jet $\sigma_{\text{dec}}$. This is based on the 'thin shell approximation' (e.g., \citealp{Hascoet2014}), where the majority of the ejecta kinetic energy is transferred to the blast wave only after sufficient ambient gas has been swept up by the external shock. A statistical analysis done by \cite{Ghirlanda2018} showed that 80 per cent of the bursts have $t_{\text{p}}>T_{90}$, which is expected in the thin shell limit, so the assumptions behind the calculation of $\Gamma_{\text{dec}}$ seem to be valid.

\section{Conclusions}

In this work, we use GRB afterglow data from the literature (the magnetization of the jet $\sigma_{\text{dec}}$ and the asymptotic Lorentz factor $\Gamma_{\text{dec}}$) to connect the afterglow region with that of the prompt emission. 
We start from a region far from the GRB central engine (the deceleration radius) and our goal is to move to smaller distances, all the way to the central engine, and gather information about its nature
and the energetics of the jet that it produces, such as the peak energy and the photospheric efficiency of the prompt emission of the burst. 

The inferred asymptotic Lorentz factors and jet luminosities imply that most of the jets in our samples are characterized by $R_{\rm diss}/R_{\rm ph}\sim 1$, i.e., naturally predicting a powerful photospheric emission component (see, e.g., Fig. \ref{fig:radiiratio}). By analyzing our sample of bursts and calculating their peak energy and photospheric efficiency, we find that a neutron star central engine is favored by our model for a majority of the bursts. The photospheric efficiency is in the $5-10$ per cent range with the maximum value depending weakly on the model's parameters $\alpha, R_0, s$. On the other hand, the peak energy of the prompt emission has a stronger parametric dependence, based on which we deduce that neutron stars are a preferable engine for our model.

From our G-Sample analysis, we also find that the model favors small values of the stripe parameter $\alpha$, which corresponds to a broad distribution of stripes that extends well above the minimum spatial scales $l_{\text{min}}$. For future work, it would be interesting to extend our sample with more bursts that have sufficient afterglow observations; doing so would allow us to explore whether the trend of small $\alpha$ values persists.

\section*{Acknowledgements}
MD and DG acknowledge support from the Fermi Cycle 14 Guest Investigator Program 80NSSC21K1951, 80NSSC21K1938, and the NSF AST-2107802 and AST-2107806 grants. RBD acknowledges support from the National Science Foundation under grants 1816694 and 2107932. 

\section*{Data Availability}

The data underlying this article will be shared on reasonable request to the corresponding author




\bibliographystyle{mnras}
\bibliography{bib} 







\bsp	
\label{lastpage}
\end{document}